%% file: FM-CchiQM-Nc.tex
\documentclass[11pt,cite]{article}
\usepackage{jheppub}

\usepackage{latexsym}

\usepackage[latin1]{inputenc}

\usepackage{color,pst-plot}

\usepackage{amsmath}
\usepackage {amsfonts,amssymb,amstext}

\newcommand{\lbl}[1]{\label{eq:#1}}
\newcommand{ \rf}[1]{(\ref{eq:#1})}

\newcommand{\be}{\begin{equation}}
\newcommand{\ee}{\end{equation}}
\newcommand{\bea}{\begin{eqnarray}}
\newcommand{\eea}{\end{eqnarray}}
\newcommand{\setl}{\setlength\arraycolsep{2pt}}

\newcommand{\noi}{\noindent}
\newcommand{\nn}{\nonumber}
\newcommand{\ra}{\rightarrow}
\newcommand{\Ra}{\Rightarrow}

\newcommand{\cL}{{\cal L}}
\newcommand{\cM}{{\cal M}}

\newcommand{\cO}{{\cal O}}

\newcommand{\Imm}{\mbox{\rm Im }}
\newcommand{\Ree}{\mbox{\rm Re }}

\newcommand{\tr}{\mbox{\rm tr}}

\newcommand{\MeV}{\mbox{\rm MeV}}
\newcommand{\GeV}{\mbox{\rm GeV}}

\newcommand{\annd}{\mbox{\rm and}}
\newcommand{\foor}{\mbox{\rm for}}

\newcommand{\Nc}{\mbox{${\rm N_c}$}}
\newcommand{\Ncb}{\mbox{${\rm\bf N_c}$}}

\input epsf

\renewcommand{\thesection}{\normalsize \Roman {section}}
\def\theequation{\arabic{section}.\arabic{equation}}

\title{Asymptotic Behaviour of Pion--Pion Total Cross--Sections}
\author[a]{David Greynat,}
\author[b]{Eduardo de Rafael}
\author[c]{and Gr\'egory Vulvert}

\affiliation[a]{Dipartimento di Scienze Fisiche, Universita  di Napoli ``Federico II''\\
Via Cintia, 80126 Napoli, Italia}
\affiliation[b]{Aix-Marseille Universit\'e, CNRS, CPT, UMR 7332, 13288 Marseille, France\\
Université de Toulon, CNRS, CPT, UMR 7332, 83957 La Garde, France}
\affiliation[c]{Departament de F\'{i}sica Te\'{o}rica, IFIC, CSIC -- Universitat de Val\`{e}ncia, \\
Apt. Correus 22085, E-46071 Val\`{e}ncia, Spain}

\emailAdd{david.greynat@gmail.com}
\emailAdd{EdeR@cpt.univ-mrs.fr}
\emailAdd{vulvert@ific.uv.es}

\abstract{We derive a sum rule which shows that the Froissart--Martin bound for the asymptotic behaviour of the $\pi\pi$ total cross sections at high energies, if modulated by the Lukaszuk--Martin coefficient of the leading $\log^2 s$ behaviour, cannot be an optimal bound in QCD. We next compute the total cross sections for $\pi^+ \pi^-$, $\pi^{\pm} \pi^0$ and $\pi^0 \pi^0$ scattering within  the framework of the  constituent chiral quark model (C$\chi$QM) in the limit of a large number of colours $\Nc$ and discuss their asymptotic behaviours. The same $\pi\pi$ cross sections are also discussed within the general framework of  Large--\Nc~QCD and we show that it is possible to make an Ansatz for the isospin $I=1$ and $I=0$ spectrum  which satisfy the Froissart--Martin bound with coefficients which, contrary to the Lukaszuk--Martin coefficient, are not singular in the chiral limit and have the correct Large--\Nc~counting. We finally propose a simple  phenomenological model which matches the low energy behaviours of the $\sigma_{\pi^{\pm}\pi^0}^{\rm total}(s)$ cross section predicted by the C$\chi$QM  with the high energy behaviour predicted by the Large--$\Nc$ Ansatz. The magnitude of these cross sections at very high energies is of the order of those observed  for the  $pp$ and $p{\bar p}$ scattering total cross sections.}

\begin{document}
\maketitle

%
%

\section{\normalsize Introduction}\lbl{int}
\setcounter{equation}{0}
\def\theequation{\arabic{section}.\arabic{equation}}

\noi
The Froissart--Martin bound for the asymptotic behaviour of total cross sections has played a major role in the history of strong interactions. Using the Mandelstam representation, Froissart showed~\cite{Fr61} that the total cross section $\sigma_{AB}^{\rm total}(s)$  for the scattering of two hadronic particles $A$ and $B$ with center of mass energy $\sqrt{s}$  cannot grow faster than
\be\lbl{eq:frois}
\sigma_{AB}^{\rm total}(s) \underset{{s \ra \infty}}{\leq}\ C \log^2 \frac{s}{s_0}\,,
\ee
with $C$ and $s_0$ unknown constants. The rigorous proof of this bound from axiomatic quantum field theory was later shown by Martin~\cite{Mart66} and, quite remarkably, an explicit form for the coefficient $C$ was first derived by Lukaszuk and Martin in ref.~\cite{LuMart67} with the result
\be
C=\frac{4\pi}{t_0}\,,
\ee
where $t_0$ denotes the lowest mass squared singularity in the $t$--channel of the scattering process\footnote{With some assumptions, the normalization $s_0$ for {\it averaged} total cross sections  has  also been recently derived in ref.~\cite{MR13}.}. 
When applied to pion--pion scattering the Froissart--Martin--Lukaszuk bound (FML--bound) states that
\be\lbl{eq:FM}
\sigma^{\mbox{\footnotesize tot}}_{\pi\pi}(s) \underset{{s \ra \infty}}{\leq}\ \frac{\pi}{m_{\pi}^2}\log^2 \frac{s}{s_0}\,.
\ee

The presence of a factor $1/m_{\pi}^2$ in the r.h.s. of  eq.~\rf{eq:FM} has  recently been questioned by two of us~\cite{GdeR13}: {\it what happens to the FML--bound  in QCD in the chiral limit when the pions, the Nambu--Goldstone states of the chiral--$SU(2)$ flavour symmetry of QCD, become massless?} Does the FML--bound become irrelevant in this limit as eq.~\rf{eq:FM} seems to indicate?
As pointed out in ref.~\cite{GdeR13}, the usual derivation of the FML--bound from the  rigorous principles of axiomatic quantum field theory does not take into account  the fact that the underlying dynamics of the strong interactions has the property of spontaneous chiral symmetry breaking. In fact, it implicitly assumes a realization  of the hadronic spectrum {\it \`a la} Wigner--Weyl without Nambu--Goldstone particles, in which case, the coefficient of the $\log^2 s$ in eq.~\rf{eq:FM} is perhaps not so surprising; but what is then the correct coefficient in QCD? 
In the next section we derive a sum rule which clearly shows that the FML--bound cannot be the optimal bound in QCD.

Another question which was also raised in ref.~\cite{GdeR13} is:
{\it what becomes of the FML--bound in the QCD Large--$N_c$~limit?} The Large--$\Nc$ counting rules fix $\sigma^{\mbox{\footnotesize tot}}_{\pi\pi}(s)$ to be of $\cO\left(1/\Nc\right)$, while the r.h.s. of eq.~\rf{eq:FM} appears to be of  $\cO(1)$.  We propose here, as a very modest step towards an answer to these fundamental  questions, to examine them first in the simple Constituent Chiral Quark Model~\cite{MG84} (C$\chi$QM) and then within the  more general framework of  Large--$\Nc$ QCD properties. 

We have organized this paper in the following way. In the next section we reproduce basic properties of the elastic pion--pion scattering amplitudes which are necessary to derive the sum rule which we have  mentioned above. 
The effective Lagrangian of the C$\chi$QM is described in section  III as well as the results of the calculation of the total $\pi\pi$ annihilation cross sections in this model. We also show that the results obtained for these cross sections satisfy the sum rules which, within the model, fix some of the  $\cO(p^4)$ low--energy constants of the chiral Lagrangian. Section IV is then dedicated to the pion--pion total cross sections in the QCD Large--$\Nc$ limit where we reconsider in more detail some of the issues discussed in ref.~\cite{GdeR13} and where we propose a model which matches the low energy behaviours of the $\sigma_{\pi\pi}^{\rm total}(s)$ cross sections predicted by the C$\chi$QM  with the high energy behaviours predicted by a simple Large--$\Nc$ Ansatz.  Phenomenological comments and conclusions are summarized in Section V.

\section{\normalsize 
Pion-Pion Amplitudes and Sum Rules}\lbl{sumrules}
\setcounter{equation}{0}
\def\theequation{\arabic{section}.\arabic{equation}}

\noi
In full generality, elastic $\pi\pi$ scattering in  the isospin symmetry limit is  described by a single invariant Lorentz amplitude $A(s,t,u)$~\footnote{For a modern review see ref.~{\cite{ACGL01}}.}.

{\setl
\bea
\lefteqn{
\langle \pi^{d}(p_4)\pi^{c}(p_3)\ {\rm out}\vert \pi^{a}(p_1) \pi^{b}(p_2)\ {\rm in}\rangle =}\nn \\
 & & \hspace*{-1cm} {\bf 1}+i(2\pi)^4 \delta^4 (p_3 +p_4 -p_1 -p_2)\left\{ \delta^{ab}\delta^{cd}A(s,t,u) + \delta^{ac}\delta^{bd}A(t,u,s)+\delta^{ad}\delta^{bc}A(u,s,t)\right\}\,,\lbl{eq:amp}
\eea}

\noi
where $a$, $b$, $c$, $d$ denote the 1,2,3 components of the adjoint representation of the pion fields in $SU(2)$ and $s$, $t$ and $u$ the usual Mandelstam variables  constrained by
\be
s+t+u=4m_{\pi}^2\,.
\ee
It is convenient to work with the three $s$--channel isospin components ${\bf T}=(T^0, T^1, T^2)$ of the amplitudes in eq.~\rf{eq:amp}:

{\setl
\bea
T^0 (s,t) & = & 3 A(s,t,u)+A(t,u,s)+A(u,s,t)\,,\nn\\
T^1 (s,t) & = & A(t,u,s)-A(u,s,t)\,,\nn\\
T^2 (s,t) & = & A(t,u,s)+A(u,s,t)\,.
\eea}

\noi
These isospin amplitudes obey fixed-$t$ dispersion relations. They are the so--called Roy equations~{\cite{Roy71}} which we shall consider at $t=0$ and in the chiral limit where $m_{\pi}\ra 0$. The linear combination of the isospin amplitudes  which diagonalize the crossing matrix in the Roy equations are then:

{\setl
\bea\lbl{eq:fsandts}
F_{1}(s,0) & = & -\frac{1}{6}\ T^0 (s,0) -\frac{1}{4}\ T^1 (s,0)+\frac{5}{12}\ T^2 (s,0) \,,\nn\\
F_{2}(s,0) & = & +\frac{1}{6}\ T^0 (s,0) +\frac{1}{4}\ T^1 (s,0)+\frac{7}{12}\ T^2 (s,0)  \,,\nn\\
F_{3}(s,0) & = & -\frac{1}{6}\ T^0 (s,0) +\frac{3}{4}\ T^1 (s,0)+\frac{5}{12}\ T^2 (s,0)  \,.
\eea}

\noi
The amplitudes $F_2$ and $F_3$  obey the same dispersion relation:
\be\lbl{eq:F23}
\Ree F_{2,3}(s,0) = s^2 \int_0^\infty \frac{\mathrm{d}s'^2}{s'^2}\frac{1}{s'^2 -s^2}\frac{1}{\pi}\Imm F_{2,3}(s',0)\,,
\ee
and $\Ree F_{2,3}(s,0)$  are even under $s\leftrightarrow -s$, while the amplitude $F_{1}(s,0)$ obeys the dispersion relation:
\be\lbl{eq:f1}
\Ree F_1 (s,0)= -\frac{s}{f_{\pi}^2}+2 s^3 \int_0 ^\infty \frac{\mathrm{d}s'}{s'^2}\frac{1}{s'^2 -s^2}\frac{1}{\pi}\Imm F_1 (s',0)\,,
\ee
and $\Ree F_1 (s,0)$ is odd under $s\leftrightarrow -s$. Indeed, one can check that there is no contribution of $\cO(s^2)$ to  $\Ree F_1 (s,0)$ in Chiral Perturbation Theory ($\chi$PT), while the contributions of that order from $\chi$PT to the $F_{2}(s,0)$ and $F_{3}(s,0)$ amplitudes are~\cite{GL84}:

{\setl
\bea
\Ree F_2 (s,0) & \underset{s \rightarrow 0}{=}\ & \frac{s^2}{f_{\pi}^4}\left[ 2l_1^{\rm r}(\mu) +3l_2^{\rm r}(\mu) +\frac{1}{12\pi^2}\left(\log\frac{\mu^2}{s}+\frac{25}{24}\right)\right] + \cO(s^4 )\,, \lbl{eq:chip42} \\
\Ree F_3 (s,0) & \underset{s \rightarrow 0}{=}\ & \frac{s^2}{f_{\pi}^4}\left[ -2l_1^{\rm r}(\mu) +l_2^{\rm r}(\mu) +\frac{1}{96\pi^2}\right] + \cO(s^4 )\,. \lbl{eq:chip43}
\eea}

\noi
These  $\chi$PT one--loop results are renormalization $\mu$--scale independent and well defined in the chiral limit. The $\mu$--scale dependence of the chiral log cancels with the $\mu$--dependence in the renormalized constants and the combination $-2l_1^{\rm r}(\mu) +l_2^{\rm r}(\mu)$ is $\mu$--scale independent.  The relation between the chiral $SU(2)$  $l_{i}^{\text r}$ constants and the more conventional $L_{i}^{\text r}$ constants of the  chiral $SU(3)$ Lagrangian is as follows~\cite{GL85}:

{\setl
\bea\lbl{eq:Llrels}
l_{1}^{\text r}(\mu) & = & 4 L_{1}^{\text r}(\mu) +2 L_{3}  -\frac{1}{96\pi^2}\frac{1}{8}\left(\log\frac{M_{K}^2}{\mu^2}+1 \right)\,,\lbl{eq:Llrels1} \\
l_{2}^{\text r}(\mu) & = & 4 L_{2}^{\text r}(\mu)  -\frac{1}{96\pi^2}\frac{1}{4}\left(\log\frac{M_{K}^2}{\mu^2}+1 \right)\,,\lbl{eq:Llrels2}
\eea}

\noi
where here, kaon particles have been treated as massive and  integrated out, hence the dependence on their mass $M_K$.

The optical theorem relates the amplitudes $\Imm F_{i}(s,0)$ to the total $\pi\pi$ cross sections as follows (massless pions):

{\setl
\bea
\Imm F_{1} (s,0)  & = & \frac{1}{2}\left[s\ \sigma_{\pi^{+}\pi^{+}}^\text{tot} -s\ \sigma_{\pi^{+}\pi^{-}}^\text{tot}\right]\,,\nn \\
\Imm F_{2} (s,0) & = & \frac{1}{2}\left[s\ \sigma_{\pi^{+}\pi^{+}}^\text{tot} +s\ \sigma_{\pi^{+}\pi^{-}}^\text{tot}\right]= \frac{1}{2}\left[s\ \sigma_{\pi^{\pm}\pi^{0}}^\text{tot} +s\ \sigma_{\pi^{0}\pi^{0
}}^\text{tot}\right]\,,\nn \\
\Imm F_{3} (s,0)& = & \frac{1}{2}\left[3s\ \sigma_{\pi^{\pm}\pi^{0}}^\text{tot} -s\ \sigma_{\pi^{0}\pi^{0}}^\text{tot}\right]\,.
\eea}

\noi

\vspace*{0.5cm} 
\subsection{\normalsize 
The Lukaszuk--Martin Bound}\lbl{martin}

\noi
We are now in the position to explain what is the problem with the Lukaszuk--Martin coefficient~\cite{LuMart67} of the FML--bound in eq.~\rf{eq:FM} as well as with the recent averaged bound derived in ref.~\cite{MR13}. For that, let us consider the sum of eqs.~\rf{eq:F23} at small $s$, say $s=m_{\pi}^2$ to be precise,  which  we rewrite in the form of a convenient sum rule: 
\be\lbl{eq:SMRR}
\frac{\pi}{f_{\pi}^4}\left[ l_2^{\rm r}(\mu) +\frac{1}{48\pi^2}\left(\log\frac{\mu^2}{m_{\pi}^2}+\frac{7}{6}\right)\right]+\cO\left(m_{\pi}^4 \right)= \int_{4 m_{\pi}^2} ^\infty \frac{ \mathrm{d}s'}{s'^2 -m_{\pi}^4}\ \sigma^{\mbox{\footnotesize tot}}_{\pi^{\pm} \pi^0}(s')\,,
\ee
where on the l.h.s. we have used eqs.~\rf{eq:chip42},~\rf{eq:chip43} and on the r.h.s. the fact that
\be
\Imm F_2 (s',0)+\Imm F_3 (s',0)= 2 s'\ \sigma^{\mbox{\footnotesize tot}}_{\pi^{\pm} \pi^0}(s')\,.
\ee
We recall that the l.h.s. does not depend on the choice of the renormalization scale $\mu$. Using the FML--bound we can write the integral in the r.h.s. as follows:

{\setl
\bea
 \int_{4 m_{\pi}^2} ^\infty \frac{\mathrm{d}s'}{s'^2 -m_{\pi}^4}\ \sigma^{\mbox{\footnotesize tot}}_{\pi^{\pm} \pi^0}(s')& \le &
\int_{4 m_{\pi}^2} ^{s_0 }\frac{\mathrm{d}s'}{s'^2 -m_{\pi}^4}\ \sigma^{\mbox{\footnotesize tot}}_{\pi^{\pm} \pi^0}(s')\nn \\
 &  &  + \, \frac{\pi}{m_{\pi}^2}\frac{2}{s_0}\left[ 1+\frac{1}{27}\frac{m_{\pi}^4}{s_0^2}+\cO\left(\frac{m_{\pi}^4}{s_0^2}\right)^2\right]\,,
\eea}

\noi
where $s_0$ is the finite threshold where the asymptotic behaviour sets in, the same $s_0$ which appears as the normalization of the $\log^2 s$ in the FML bound in Eq.~\rf{eq:FM}.
Both terms in the r.h.s. of this inequality are obviously positive. 
Inserting this inequality in the r.h.s. of the sum rule in eq.~\rf{eq:SMRR} shows that the l.h.s., which goes as $\log m_{\pi}^2$ in the chiral limit, is bounded by a quantity which diverges as $1/{ m_{\pi}^2}$  in the same limit. We therefore conclude that, if the Froissart  bound applies to $\pi\pi$ scattering in QCD, the Lukaszuk--Martin coefficient  of the leading $\log^2 s$ term cannot be the optimal one.

We wish to emphasize that the derivation of this result follows from very general properties of $\pi\pi$ amplitudes in QCD and, so far, we have not consider the Large--$\Nc$ approximation.   

\vspace*{1.2cm} 
\section{\normalsize The Constituent Chiral Quark Model (C$\chi$QM)}\lbl{CQM}
\setcounter{equation}{0}
\def\theequation{\arabic{section}.\arabic{equation}}

\noi
Historically, the model in question emerged as an attempt to reconcile  the  successes of phenomenological quark models, like e.g. the De~R\'ujula-Georgi-Glashow model~\cite{deRGG75}, with  QCD. The Lagrangian proposed by Manohar and Georgi (MG) is an effective field theory which incorporates the interactions of the low--lying pseudoscalar particles of the hadronic spectrum i.e., the Nambu-Goldstone modes of the spontaneously broken chiral symmetry~\cite{Wei79}, with chirally rotated quark fields $Q=(U,D,S)$. These quarks  have become massive due to the phenomenon of  spontaneous chiral symmetry breaking (S$\chi$SB). Their mass, however, ($M_Q \sim \frac{1}{3}M_{\rm nucleon}$) has nothing to do with the masses of the $u,d,s$ quark fields in the QCD Lagrangian which explicitly break chiral symmetry and are known to be much smaller (see e.g. the recent review article in ref.~\cite{FLAG13}).  The constituent quark fields may also have gluonic interactions but, since the Goldstone modes are already in the Lagrangian, the color--${\rm SU(3)}$ coupling constant  is then supposed to be no longer running and relatively small. The hope is that such an effective Lagrangian encodes the essential degrees of freedom to describe Hadron Physics at energies below the chiral symmetry breaking scale but still above the confinement regime.
 
It is fair to say, however, that in spite of some efforts (see e.g. refs.~\cite{EdeRT90,Wei10,Kapl13} and references therein), it has not been possible to establish the approximations at which the MG--Lagrangian could be derived from the underlying QCD theory. It can be shown to be a particular case of the Extended Nambu Jona-Lasinio (ENJL) Model~\cite{NJL61,BBdeR93}, but this only transfers the problem of its derivation from first principles to  another level where, in any case, the basic question  remains so far unanswered.

An interesting observation made by Weinberg~\cite{Wei10} is the fact that in the limit of a large number of colors $\Nc$ the C$\chi$QM becomes a renormalizable theory. A subsequent observation along the same line was made in ref.~\cite{deR11} where it is shown that the  number of counterterms which in the Large--\Nc~limit  have to be added to the primitive Manohar--Georgi Lagrangian, is minimized for the choice $g_A =1$. In that respect, it has also been shown~\cite{deR11,GdeR12} that there is a class of observables governed by integrals of  specific QCD Green's functions  which, for $g_A =1$, have rather good matching to their short--distance behaviour. Interesting examples are the Hadronic Vacuum Polarization, the Hadronic Light--by--Light Scattering and the Hadronic Electroweak contributions to the anomalous magnetic moment of the muon, which have been recently discussed within the framework of the C$\chi$QM in ref.~\cite{GdeR12}.

The  effective Lagrangian in question is the following:

{\setl
\bea\lbl{CCQL}
\cL_{{\rm C}\chi{\rm QM}}(x) & \!\! =\!\! & \underbrace{i{\bar Q}\gamma^{\mu}\left(\partial_{\mu}+\Gamma_{\mu}+iG_{\mu} \right)Q+\frac{i}{2}{g_A}\ {\bar Q}\gamma^{\mu}\gamma_5 \xi_{\mu}Q-M_{Q} {\bar Q}Q}_{\it M-G}-\frac{1}{2}{\bar Q}\left(\Sigma -\gamma_5 \Delta \ \right)Q \nn\\ 
 &  & + \underbrace{\frac{1}{4}{ f_{\pi}}^2 \tr\left[ D_{\mu}UD^{\mu}U^{\dagger}\right.}_{\it M-G}
+\left. U^{\dagger}\chi+\chi^{\dagger}U\right]-\underbrace{\frac{1}{4}\sum_{a=1}^{8} G_{\mu\nu}^{(a)}G^{(a)\mu\nu}}_{\it M-G}+ e^2 { C}\ \tr (Q_R U Q_L U^{\dagger}) \nn \\
 &  & + 
{ L_5}\  \tr D_{\mu}U^{\dagger}D^{\mu}U(\chi^{\dagger}U+U^{\dagger}\chi) +
{ L_8}\  \tr (U\chi{\dagger}U\chi{\dagger}+U^{\dagger}\chi U^{\dagger}\chi)\,.
\eea}

\noi
The underbraced terms are  those of the MG--Lagrangian, but in the presence
of external $SU(3)$ vector $v_{\mu}(x)$ and axial-vector $a_{\mu}(x)$ sources. The field matrix $U(x)$ denotes the  3$\times$3 unitary matrix in the flavour space which collects the Nambu-Goldstone fields and which under chiral rotations transforms as $U\ra V_R U V_L^{\dagger}$. The vector field matrix $D_{\mu}U$ is  the  covariant derivative of $U$:
\be
D_{\mu}U=\partial_{\mu}U-ir_{\mu}U+iUl_{\mu}\,,\quad l_{\mu}=v_{\mu}-a_{\mu}\,,\quad r_{\mu}=v_{\mu}+a_{\mu}\,, 
\ee 
and, with $U=\xi\xi$,
\be
\Gamma_{\mu}=\frac{1}{2}\left[
\xi^{\dagger}(\partial_{\mu}-ir_{\mu})\xi+
\xi(\partial_{\mu}-il_{\mu})\xi^{\dagger}\right]\,,\quad
\xi_{\mu}=i\left[\xi^{\dagger}(\partial_{\mu}-ir_{\mu})\xi-
\xi(\partial_{\mu}-il_{\mu})\xi^{\dagger}\right]\,.
\ee
The gluon field matrix in the fundamental representation of color $SU(3)$ is $G_{\mu}(x)$ and $G_{\mu\nu}^{(a)}(x)$ its corresponding gluon field strength tensor.
The presence of external scalar $s(x)$ and pseudoscalar $p(x)$ sources  induces the extra terms proportional to
\be
\chi= 2 B [s(x)+i p(x)]\,,
\ee
where $B$, like $f_{\pi}$, are order parameters not fixed by the model. When the $s(x)$ and $p(x)$  sources are frozen to the up, down, and strange light quark masses of the QCD Lagrangian, 
\be
\chi= 2 B \cM\,,\quad {\rm with}\quad \cM={\rm diag}(m_u\,, m_d\,, m_s)\,,
\ee
and then
\be
\Sigma=\xi^{\dagger}\cM \xi^{\dagger}+\xi\cM^{\dagger} \xi\,,\quad 
\Delta=\xi^{\dagger}\cM \xi^{\dagger}-\xi\cM^{\dagger}\xi\,.
\ee

With the axial coupling  fixed to  $g_A =1$,
the extra couplings $L_5$ and $L_8$
are the only terms which are needed to absorb the ultraviolet (UV) divergences  when the constituent quark fields $Q(x)$ are integrated out\footnote{We disregard couplings involving external fields alone to lowest order in the chiral expansion.}. If one wants to consider the case where photons are also integrated out then, to leading order in the chiral expansion and in the electric charge coupling $e$, the last term in the second line is also required to absorb further UV--divergences. 
Loops involving pion fields are subleading in the $1/{\rm N_c}$--expansion and hence, following the observation of Weinberg in ref.~\cite{Wei10}, the Lagrangian in eq.~\rf{CCQL}, when considered within the framework of the large--${\rm N_c}$ limit, is a renormalizable Lagrangian.

The  $\pi\pi$ total cross sections in the C$\chi$QM are then simply given by the corresponding
$
\pi\pi\ra Q\bar{Q}
$
cross sections\footnote{For the sake of simplicity we omit in this first analysis the contribution of gluon interactions in the C$\chi$QM.}. When restricted to chiral $SU(2)$  and in the chiral limit where $m_u =m_d =0$, but still keeping the value of $g_A$ free, the terms of the interaction Lagrangian which  are needed for the evaluation of these  cross sections are [$Q=(U,D)$]
:

{\setl
\bea\lbl{eq:intlagr}
\cL_{\rm int}(x) & \doteq & -i  \frac{g_A}{\sqrt{2} f_{\pi}}\left[\bar{U}\gamma^{\mu}\gamma_{5}D\  \partial_{\mu}\pi^+ + \bar{D}\gamma^{\mu}\gamma_{5}U\ \partial_{\mu}\pi^-  +\frac{1}{\sqrt{2}}\left( \bar{U}\gamma^{\mu}\gamma_{5}U-\bar{D}\gamma^{\mu}\gamma_{5}D\right)\ \partial_{\mu}\pi^0  \right] \nn \\
 & & +i\frac{1}{4 f_{\pi}^2}\left\{ \bar{U}\gamma^{\mu}U\ \left(\pi^+ \partial_{\mu}\pi^- - \pi^- \partial_{\mu}\pi^+ \right)+\bar{D}\gamma^{\mu}D\  \left(\pi^- \partial_{\mu}\pi^+ - \pi^+ \partial_{\mu}\pi^- \right)\right\}\nn\\
 &  & 
+i\frac{\sqrt{2}}{4 f_{\pi}^2}\left\{\bar{U}\gamma^{\mu}D\  (\pi^0 \partial_{\mu}\pi^+ - \pi^+ \partial_{\mu} \pi^0 ) -\bar{D}\gamma^{\mu}U\  (\pi^0 \partial_{\mu}\pi^- - \pi^- \partial_{\mu}\pi^0 )\right\}\nn \\
 & & -M_Q \sum_{Q=U,D} \bar{Q}Q\,.
\eea}

\vspace*{1.2cm} 
\subsection{\normalsize Calculation of the  Pion-Pion Total Cross Sections}\lbl{calculation}
\def\theequation{\arabic{section}.\arabic{equation}}

\noi 
We fix the kinematics as follows:
\be 
\pi(k)+\pi (k')\ra Q(p)+ \bar{Q}(p')\,.
\ee
In the center of mass system:

{\setl
\bea\lbl{eq:lagr}
k~: & \left(\frac{\sqrt s}{2}\,,  +\frac{\sqrt s}{2}\,,  0\,,  0\right)\,,\quad p~: & \left(\frac{\sqrt s}{2}\,,  +\frac{\sqrt {s-4M_Q^2}}{2}\cos\theta\,,  
+\frac{\sqrt {s-4M_Q^2}}{2}\sin\theta\,,  0\right)\,; \nn \\
k': & \left(\frac{\sqrt s}{2}\,,  -\frac{\sqrt s}{2}\,,   0\,,  0\right)\,, \quad p': & \left(\frac{\sqrt s}{2}\,,   -\frac{\sqrt{s-4M_Q^2}}{2}\cos\theta\,,  
-\frac{\sqrt {s-4M_Q^2}}{2}\sin\theta\,,   0\right) \,.
\eea}

\noi
The  total cross sections for massless pions are then  given by
\be
\sigma^{\mbox{\footnotesize tot}}_{\pi\pi}(s)=\frac{1}{32\pi s}\sqrt{1-\frac{4M_Q^2}{s}}\int_{-1}^{+1} \mathrm{d}(\cos\theta)\left\vert T(\pi\pi\ra Q\bar{Q})\right\vert^2 (s,\cos\theta)\,,
\ee
with $\left\vert T(\pi\pi\ra Q\bar{Q})\right\vert^2 (s,\cos\theta)$ receiving contributions from the relevant terms shown in the interaction Lagrangian in eq.~\rf{eq:intlagr}.

\subsubsection{\normalsize The  $\pi^+ \pi^-$ Total Cross Section}\lbl{pipluspiminus}
\noi
The various  contributions to this cross section come from the following sources:
\begin{itemize}
	\item {Terms proportional to $\frac{g_A^4}{f_{\pi}^4} N_c n_f$.}
	
	They come from the squared amplitudes $\pi^+ \pi^- \ra U\bar{U}$ and $\pi^+ \pi^- \ra D\bar{D}$  generated by the  terms  in the first line in eq.~\rf{eq:intlagr}. They give a contribution 
	\be
\frac{g_A^4}{f_{\pi}^4} N_c n_f \Ra  \frac{N_c n_f}{192\pi }\frac{g_{A}^4}{ f_{\pi}^4}\sqrt{1-\frac{4M_Q^2}{s}}\left(s + 2 M_Q^2 \right) \,.
\ee
	
	\item {Terms proportional to $\frac{1}{f_{\pi}^4}N_c n_f$.}
	
	They come from the squared amplitudes $\pi^+ \pi^- \ra U\bar{U}$ and $\pi^+ \pi^- \ra D\bar{D}$ generated by the terms in the second line of eq.~\rf{eq:intlagr}. They give a contribution
	\be
\frac{1}{f_{\pi}^4}N_c n_f \Ra  \frac{N_c n_f}{192\pi}\frac{1}{ f_{\pi}^4}\sqrt{1-\frac{4M_Q^2}{s}}\left(s+2M_Q^2 \right)\,.
\ee
	
	\item {Terms proportional to $\frac{g_A^2}{f_{\pi}^4}N_c n_f$}
	
	They come from the  the interference of the amplitudes $\pi^+ \pi^- \ra U\bar{U}$ and $\pi^+ \pi^- \ra D\bar{D}$ in the first and second lines of eq.~\rf{eq:intlagr}. They give a contribution
	\be
 \frac{g_A^2}{f_{\pi}^4}N_c n_f \Ra  
	-\frac{N_c n_f}{192\pi}\frac{g_A^2}{f_{\pi}^4}\sqrt{1-\frac{4M_Q^2}{s}}
2\left(s-10M_{Q}^2\right) \,.
\ee
\end{itemize}

The overall contribution to the $\pi^+ \pi^-$ total cross section is then
\begin{equation}\lbl{eq:totalpm}
\sigma^{\mbox{\footnotesize tot}}_{\pi^+ \pi^-}(s)=
\frac{N_c n_f}{192\pi}\frac{M_Q^2}{ f_{\pi}^4}\sqrt{1-\frac{4M_Q^2}{s}}\left[(1-g_A^2)^2\ \frac{s}{M_Q^2}+2(1+10g_A^2 +g_A^4)
\right] \,.	
\end{equation}
We therefore find that there is a  leading contribution of $\cO(s)$	to this cross section which, however,  vanishes for $g_A =1$.
In other words, 
the  $\pi^+ \pi^-$ scattering total cross section in the C$\chi$QM violates the Froissart bound unless  $g_A = 1$.

Equation~\rf{eq:totalpm}, for $g_A =1$,  reduces to
\be\lbl{eq:finalpm}
\sigma^{\mbox{\footnotesize tot}}_{\pi^+ \pi^-}(s)
=\frac{N_c n_f}{32\pi}\frac{4M_Q^2}{ f_{\pi}^4}\sqrt{1-\frac{4M_Q^2}{s}}\,,
\ee
and, therefore, for large--$s$:
\be\lbl{eq:aspm}
\sigma^{\mbox{\footnotesize tot}}_{\pi^+ \pi^-}(s) \underset{{s \ra \infty}}{\thicksim}
\frac{N_c n_f}{32\pi}\frac{4M_Q^2}{ f_{\pi}^4}\,.
\ee

We then conclude that the leading contribution to the  $\pi^+ \pi^-$ scattering total cross section for large $s$, in the C$\chi$QM and provided $g_A =1$ with neglect of gluon corrections, goes to a constant. There is no factor $1/m_{\pi}^2$ in this bound and it satisfies the Large--\Nc~counting rules; but it does not saturate the FM--bound.

\subsubsection{\normalsize The  $\pi^\pm \pi^0$ Total Cross Sections}\lbl{pipluspizero}
\noi
The  various  contributions to $\sigma^{\mbox{\footnotesize tot}}_{\pi^{\pm} \pi^0}(s)$  come from the following sources:
\begin{itemize}
	\item {Terms proportional to $\frac{g_A^4}{f_{\pi}^4} N_c n_f$.}
	
	They come from the squared amplitudes  generated by the  terms  in the first line in eq.~\rf{eq:intlagr}. They give a contribution 
	\be
\frac{g_A^4}{f_{\pi}^4} N_c n_f \Ra  \frac{N_c n_f}{192\pi }\frac{g_{A}^4}{ f_{\pi}^4}\left\{\sqrt{1-\frac{4M_Q^2}{s}}\left(s -11 M_Q^2 \right)+48\frac{M_Q^4}{s}\log\frac{1+\sqrt{1-\frac{4M_Q^2}{s}}}{1-\sqrt{1-\frac{4M_Q^2}{s}}}\right\}\,.
\ee
	
	\item {Terms proportional to $\frac{1}{f_{\pi}^4}N_c n_f$.}
	
	They come from the squared amplitudes  generated by the terms in the second line of eq.~\rf{eq:intlagr}. They give a contribution
	\be
\frac{1}{f_{\pi}^4}N_c n_f \Ra  \frac{N_c n_f}{192\pi}\frac{1}{ f_{\pi}^4}\sqrt{1-\frac{4M_Q^2}{s}}\left(s-2M_Q^2 \right)\,.
\ee
	
	\item {Terms proportional to $\frac{g_A^2}{f_{\pi}^4}N_c n_f$}
	
	They come from the  the interference of  amplitudes  in the first and second lines of eq.~\rf{eq:intlagr}. They give a contribution
	\be
\frac{g_A^2}{f_{\pi}^4}N_c n_f \Ra  
	-\frac{N_c n_f}{192\pi}\frac{g_A^2}{f_{\pi}^4}\sqrt{1-\frac{4M_Q^2}{s}}
	\left(2s-11M_{Q}^2\right) \,.
\ee
\end{itemize}

The overall contribution is then

\setl
\bea\lbl{eq:totalmo}
\sigma^{\mbox{\footnotesize tot}}_{\pi^{\pm} \pi^0}(s) & = & \frac{N_c n_f}{192\pi}\frac{M_Q^2}{ f_{\pi}^4}\left\{\sqrt{1-\frac{4M_Q^2}{s}}\left[\frac{s}{M_Q^2}(1-g_A^2)\left[(1-g_A^2)-2(1+11g_A^2) \right]\right]\right. \nn \\
&  & \left. +\ 12g_A^4 \frac{4M_Q^2}{s}\log\frac{1+\sqrt{1-\frac{4M_Q^2}{s}}}{1-\sqrt{1-\frac{4M_Q^2}{s}}}\right\} \,.	
\eea

\noi
We also find that there is a  leading contributions of $\cO(s)$	to the total cross section which, like in the $\pi^+ \pi^-$ case, vanishes for $g_A =1$.
Therefore, 
the  $\pi^{\pm} \pi^0$ scattering total cross section in the C$\chi$QM also violates the Froissart bound unless  $g_A = 1$.

Equation~\rf{eq:totalmo}, for $g_A =1$,  reduces to
\be\lbl{eq:finalmo}
\sigma^{\mbox{\footnotesize tot}}_{\pi^{\pm} \pi^0}(s)
=\frac{N_c n_f}{16\pi}\frac{4M_Q^2}{ f_{\pi}^4}\frac{M_Q^2}{s}\log\frac{1+\sqrt{1-\frac{4M_Q^2}{s}}}{1-\sqrt{1-\frac{4M_Q^2}{s}}}\,,
\ee
which asymptotically, for large--$s$, goes as:
\be\lbl{eq:asymppluszero}
\sigma^{\mbox{\footnotesize tot}}_{\pi^{\pm} \pi^0}(s) \underset{{s \ra \infty}}{\thicksim}
\frac{N_c n_f}{16\pi}\frac{4M_Q^2}{ f_{\pi}^4}\ \frac{M_Q^2}{s}\ \log\frac{s}{M_Q^2}\,,
\ee
and is subleading when compared to the corresponding asymptotic behaviour of $\pi^+ \pi^-$ in eq.~\rf{eq:aspm}.

\subsubsection{\normalsize The  $\pi^0 \pi^0$ Total Cross Section}\lbl{pizeropizero}
\noi
The only term in the interaction Lagrangian in eq.~\rf{eq:intlagr} which contributes to this process is the one in the last term of the first line. Therefore, the contribution to  $\sigma^{\mbox{\footnotesize tot}}_{\pi^0 \pi^0}(s)$  only comes from:
\begin{itemize}
	\item {Terms proportional to $\frac{g_A^4}{f_{\pi}^4} N_c n_f$.}
	
\end{itemize}

The overall contribution is then

\setl
\be\lbl{eq:totaloo}
\sigma^{\mbox{\footnotesize tot}}_{\pi^0 \pi^0}(s) =  \frac{N_c n_f}{32\pi}\frac{4M_Q^2}{ f_{\pi}^4}\ g_A^4\  \left\{\sqrt{1-\frac{4M_Q^2}{s}}-2\frac{M_Q^2}{s}\log\frac{1+\sqrt{1-\frac{4M_Q^2}{s}}}{1-\sqrt{1-\frac{4M_Q^2}{s}}}\right\} \,.	
\ee
and contrary to the previous $\sigma^{\mbox{\footnotesize tot}}_{\pi^+ \pi^-}(s)$ and $\sigma^{\mbox{\footnotesize tot}}_{\pi^{\pm} \pi^0}(s)$ cases one does not need to fix $g_A =1$ to respect the Froissart bound.

Asymptotically, for large--$s$ and with $g_A =1$,  it goes as:
\be\lbl{eq:sigma00as}
\sigma^{\mbox{\footnotesize tot}}_{\pi^0 \pi^0}(s) \underset{{s \ra \infty}}{\thicksim}
\frac{N_c n_f}{32\pi}\frac{4M_Q^2}{ f_{\pi}^4}\left[1-2 \frac{M_Q^2}{s}\ \log\frac{s}{M_Q^2}\right]\,,
\ee
i.e. as a constant.

The total $\pi\pi$ cross sections calculated above, with $g_A =1$, are plotted in figure~\ref{fig:GdeRVFig1} below. 
The higher curve corresponds to $\sigma^{\rm total}_{\pi^+ \pi^-}(s)$, the middle curve to $\sigma^{\rm total}_{\pi^0 \pi^0}(s)$ and the low curve to $\sigma^{\rm total}_{\pi^{\pm} \pi^0}(s)$. The cross sections are given in millibarn units and they are plotted versus $s$, the total center of mass  squared energy, in $\GeV^2$. The constituent quark mass which we have used is the center value $M_Q = 240~\MeV$  of  the result $M_Q = (240\pm 10)~\MeV$ obtained in ref.~\cite{GdeR13}. Asymptotically, $\sigma^{\mbox{\footnotesize tot}}_{\pi^+ \pi^-}(s)$ and $\sigma^{\mbox{\footnotesize tot}}_{\pi^0 \pi^0}(s)$ have the same constant behaviour:   
\be
\sigma^{\mbox{\footnotesize tot}}_{\pi \pi}(s) \underset{{s \ra \infty}}{\thicksim}
\frac{N_c n_f}{32\pi}\frac{4M_Q^2}{ f_{\pi}^4}\,,\quad\foor\quad \pi^+ \pi^-\quad\annd\quad \pi^0\pi^0\,,
\ee 
while $\sigma^{\mbox{\footnotesize tot}}_{\pi^\pm \pi^0}(s)$ falls as the difference between  $\sigma^{\mbox{\footnotesize tot}}_{\pi^+ \pi^-}(s)$ and $\sigma^{\mbox{\footnotesize tot}}_{\pi^0 \pi^0}(s)$ (see eq.~\rf{eq:asymppluszero}) and it is subleading. We therefore find that the $\sigma^{\mbox{\footnotesize tot}}_{\pi \pi}(s)$ cross sections in the C$\chi$QM satisfy the Froissart--Martin bound, but they do not saturate it. Very likely this is a drawback of the model, which ceases to be reliable at center of mass energy values of the order $\sqrt{s}\sim 2~\GeV$. It seems plausible that the rescattering of the constituent quarks in the presence of a gluonic background may be at the origin of an asymptotic black disc like behaviour, similar to the one observed in $pp$ and $p{\bar p}$  scattering~\cite{BH12,MS13,BH13}, which provides the Froissart--Martin $\log^2 s$ enhancement.
 
For a discussion of the present experimental situation, phenomenological models and future prospects concerning  $\sigma^{\mbox{\footnotesize tot}}_{\pi \pi}(s)$ cross sections at high energies see e.g. refs.~\cite{PY04,GPSS10}.

\vspace*{1cm}


\begin{figure}[!h]

	\centering

	\resizebox{0.75\textwidth}{!}{\input{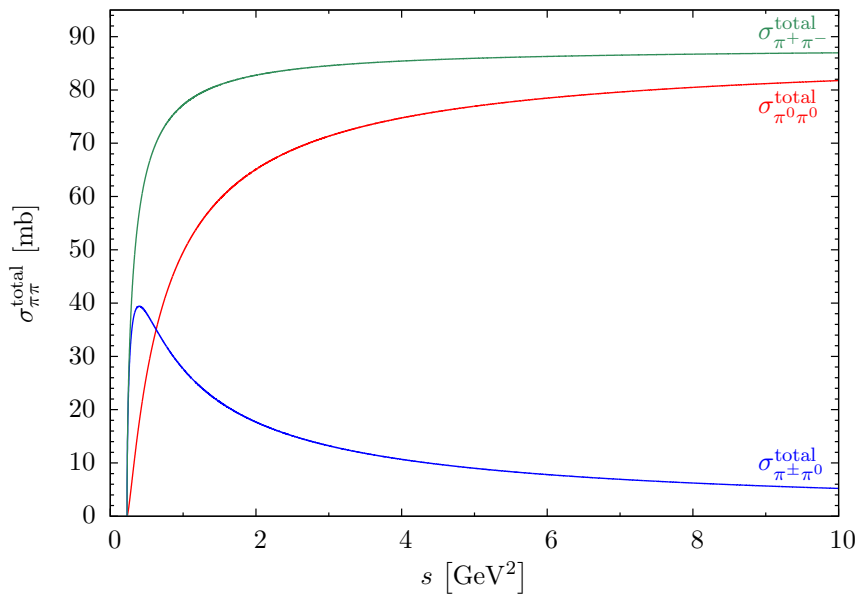}}

    \caption{Total $\pi \pi$ cross sections in millibarns versus $s$ in $\mathrm{GeV}^2$.}

    \label{fig:GdeRVFig1}

\end{figure}


\subsection{\normalsize Sum Rules in the C$\chi$QM}
\def\theequation{\arabic{section}.\arabic{equation}}

\noi

In the C$\chi$QM the total $\pi^+ \pi^+$ cross section vanishes and therefore $\Imm T^{2}(s,0)=0$, which implies:
 \be
 \Imm F_1(s,0)=-\Imm F_2 (s,0)\,,
 \ee
 and also the relations (with $g_A=1$):
 
 {\setl
 \bea\lbl{eq:im2QM}
 \Imm F_2 (s,0) & = & \frac{s}{2}\left[\sigma^{\rm tot}_{\pi^{\pm} \pi^0}+\sigma^{\rm tot}_{\pi^0 \pi^0} \right](s)=\frac{s}{2}\sigma^{\rm tot}_{\pi^+ \pi^-}(s)\\
  & = & \frac{s}{2}\frac{N_c n_f}{32\pi}\frac{4M_Q^2}{ f_{\pi}^4}  \sqrt{1-\frac{4M_Q^2}{s}}\,,
 \eea}
 
 \noi 
 and
 
 {\setl
 \bea\lbl{eq:im3QM}
 \Imm F_3 (s,0) & = & \frac{s}{2}\left[3\sigma^{\rm tot}_{\pi^{\pm}
 \pi^0}-\sigma^{\rm tot}_{\pi^0 \pi^0} \right](s)\\
  & = &  \frac{s}{2} \frac{N_c n_f}{32\pi}\frac{4M_Q^2}{ f_{\pi}^4} \left\{-\sqrt{1-\frac{4M_Q^2}{s}}+8\frac{M_Q^2}{s}\log\frac{1+\sqrt{1-\frac{4M_Q^2}{s}}}{1-\sqrt{1-\frac{4M_Q^2}{s}}}\right\}\,.
 \eea}
 
\noi 
 Inserting these expressions  in the r.h.s. of the dispersion relations in eq.~\rf{eq:F23} results in the sum rules:
\be
  \frac{2}{\pi}\int_{4 M_Q^2}^{\infty} \frac{ds'}{s'^3}\Imm F_{2,3} (s',0) =  
 \frac{1}{f_{\pi}^4}\frac{1}{3}\frac{N_c n_f}{16\pi^2}\,,
\ee
 which for $n_f =2$ reproduce the   C$\chi$QM results for $\Ree F_{2,3} (s,0)$ in eqs.~\rf{eq:chip42} and~\rf{eq:chip43}, with $g_A =1$,  and where~\cite{EdeRT90}:
 \be\lbl{eq:lcqm}
 2l_1 +3l_2 = -2l_1 +l_2 = \frac{2}{3}\frac{N_c}{16\pi^2}\,.
 \ee
 
 In fact, the chiral expansions of the $\Ree F_{1,2,3}(s,0)$ amplitudes in the C$\chi$QM, evaluated with the dispersion relations in eqs.~\rf{eq:F23}  and~\rf{eq:f1} are:
 
{\setl
\bea
\Ree F_{1}(s,0) & \underset{s \rightarrow 0}{=}\  & -\frac{s}{f_{\pi}^2}\left(1+\frac{N_c n_f}{ 240 \pi^2}\frac{s^2}{f_{\pi}^2 M_Q^2}+\cO(s^4 /M_Q^8)\right) \,, \\
\Ree F_{2}(s,0) & \underset{s \rightarrow 0}{=}\  & \frac{s^2}{f_{\pi}^4}\frac{N_c n_f}{48\pi^2}\left(1+\frac{1}{70}\frac{s^2}{M_{Q}^4}+\cO(s^4 /M_Q^8)\right)\,,\\
\Ree F_{3}(s,0) & \underset{s \rightarrow 0}{=}\  & \frac{s^2}{f_{\pi}^4}\frac{N_c n_f}{48\pi^2}\left(1+\frac{1}{35}\frac{s^2}{M_{Q}^4}+\cO(s^4 /M_Q^8)\right)\,.
\eea}

\noi 
 
\vspace*{1.2cm} 
\section{\normalsize 
Pion-Pion Cross Sections in Large-\Ncb QCD}\lbl{largeNc}
\setcounter{equation}{0}
\def\theequation{\arabic{section}.\arabic{equation}}

\noi
Let us now consider the pion--pion amplitudes within the framework  of Large--$\Nc$ QCD. 
In this limit  the $\Imm T^{I}(s,0)$ isospin amplitudes   are composed of an infinite set of narrow states:
\be\lbl{eq:largeAnsatz}
\frac{1}{\pi}\Imm T^{I}(s,0)=\sum_{n=0}^{\infty}\vert F_{I,n}\vert^2 \, \delta\left(s-M_{I,n}^2\right)\,,\quad I=0,1\,.
\ee
The spacing of the narrow states, explored in Lattice Large--$\Nc$ QCD simulations~\cite{lattice}, is compatible with the Regge growth of the leading trajectories\footnote{See e.g. ref.~\cite{AAS00}.} ($n=0,1,2,3,\dots$):
\be 
M_{I=1,n}^2 = \frac{1}{\alpha_1}\left(2n+1 -\alpha_0\right)\quad\annd\quad
M_{I=0,n}^2 = \frac{1}{\alpha_1}\left(2n+2 -\alpha_0\right)\,,
\ee
and, to a good approximation, with $\alpha_1$ and $\alpha_0$ as given by the Veneziano model~\cite{VEN68}:
\be 
\alpha_1 \simeq \frac{1}{2 M_{\rho}^2}\quad \annd\quad \alpha_0 \simeq 1/2\,.
\ee
With $M_{\rho}=770~\MeV$, this choice corresponds to:
\be
M_{\sigma}= 1334~\MeV\sim f_2 (1270)\quad\annd\quad\Lambda_{\rho}=\Lambda_{\sigma}=2 M_{\rho}\,.
\ee
In the absence of exotic trajectories and hence no poles with $I=2$, the optical theorem relates the imaginary parts of the forward isospin amplitudes to the total $\pi\pi$ cross sections as follows:

{\setl
\bea
\Imm T^1 (s,0) & = & \frac{s}{2}\left[\sigma_{\pi^+ \pi^-}^{\rm total}(s)+3 \sigma_{\pi^{\pm} \pi^0}^{\rm total}(s)-\sigma_{\pi^0 \pi^0}^{\rm  total}(s)\right]=\frac{s}{2}\ 4\ \sigma_{\pi^{\pm} \pi^0}^{\rm total}(s)\,. \lbl{eq:isocross1}\\
\Imm T^0 (s,0) & = & \frac{s}{2}\  6\ \sigma_{\pi^0 \pi^0}^{\rm  total}(s)\,.\lbl{eq:isocross0}
\eea}

\noi

Unfortunately, there is no  information on the values of the residues $\vert F_{I,n}\vert$. As shown in ref.~\cite{GdeR13}, demanding that the  cross sections  which define $\Imm T^{I=1}(s,0)$ and $\Imm T^{I=0}(s,0)$ in eqs.~\rf{eq:isocross1} and~\rf{eq:isocross0} grow asymptotically as the Froissart bound,  requires the couplings $\vert F_{I,n}\vert^2$ to grow  like $n\log^2 n$ as $n\ra \infty$. This suggests as  a possible set of  Large--$\Nc$ $\Imm T^{I} (s,0)$ amplitudes  the simple Ansatz: 

\be \lbl{eq:largeAnsatz}
\frac{1}{\pi}\Imm T^{I}(s,0)  =  {\rm C_I} \sum_{n=0}^\infty 
\left(M_{I}^2 +n\Lambda_{I}^2 \right) \log^2 \left(\frac{M_{I}^2}{\Lambda_{I}^2} +n \right)\delta(s-M_{I}^2 -n\Lambda_{I}^2 ) \,,
\ee
where:
\be \lbl{eq:parameters}
\Lambda_{I=1}  =  \Lambda_{\rho}\,,M_{I=1}=M_{\rho}\quad\annd\quad\Lambda_{I=0}=\Lambda_{\sigma}\,, M_{I=0}=M_{\sigma}\,,
\ee

\noi and the ${\rm C_I}$ are dimensionless constants. This is essentially  the same Ansatz which was considered in ref.~\cite{GdeR13}.
There is, however, an important extrapolation which one is making in assuming such an Ansatz, namely the fact that the $n\log^2 n$ pattern of the residues $\vert F_{I,n}\vert$ is valid not only at very high energies but also at low energies. Because of that extrapolation, the corresponding values that one obtains for the coefficients of the  $\log^2 s$ behaviour of the total $\sigma_{\pi\pi}^{\rm total}$ cross sections are likely to be a gross  overestimation. In what follows, we propose  a more elaborated Large--$\Nc$ Ansatz where the residues $\vert F_{I,n}\vert$ start growing like $n$, as suggested by the Veneziano model, and only from a threshold $s_0$ onwards, where asymptotics sets in, are they  modified by terms which grow as $n \log^2 n$. More precisely, we suggest the following Large--$\Nc$ Ansatz:
\be \lbl{eq:largeAnsatzimp}
\frac{1}{\pi}\Imm T^{I}(s,0)  =   \sum_{n=0}^\infty 
\left(M_{I}^2 +n\Lambda_{I}^2 \right)\left[{\rm C_I^{V}}+{\rm C_I^{F}}\ \theta(n-N)  \log^2 \left(\frac{M_{I}^2}{\Lambda_{I}^2} +n \right)\right]\delta(s-M_{I}^2 -n\Lambda_{I}^2 )\,,
\ee 
with $N$ a sufficiently large integer so as to  match the asymptotic threshold
\be
s_0 = M_I^2 +N \Lambda_I^2\,, 
\ee
and the mass parameters $\Lambda_{I=1}= \Lambda_{\rho}\,, M_{I=1}=M_{\rho}\quad\annd\quad\Lambda_{I=0}=\Lambda_{\sigma}\,, M_{I=0}=M_{\sigma}$ the same as in eq.~\rf{eq:parameters}. Equation   \rf{eq:largeAnsatzimp} can also be written in the more convenient form:

{\setl
\bea \lbl{eq:largeAnsatzimp2}
\frac{1}{\pi}\Imm T^{I}(s,0) & =  & {\rm C_I^{V}} \sum_{n=0}^\infty 
\left(M_{I}^2 +n\Lambda_{I}^2 \right)\delta(s-M_{I}^2 -n\Lambda_{I}^2 )\nn \\
&  & + \; {\rm C_I^{F}}\sum_{n=0}^\infty (s_0 +n\Lambda_{I}^2) \log^2 \left(\frac{s_0}{\Lambda_{I}^2} +n \right)\delta(s-s_0 -n\Lambda_{I}^2 )\,.
\eea}

\noi
 
The Mellin transforms of infinite sums like the ones in eq.~\rf{eq:largeAnsatzimp2} have a close analytic form:

{\setl 
\bea\lbl{eq:truemellin}
\Sigma^{I}(\xi) &  = & \int_{0}^{\infty} \mathrm{d}\left( \frac{s}{\Lambda_{I}^2}\right) \left(\frac{s}{\Lambda_{I}^2} \right)^{\xi -1} \frac{1}{\pi}\Imm T^{I} (s,0) \\
 & =  &    {\rm C_{I}^V}\ \zeta\left( -\xi,\frac{M_{I}^2}{\Lambda_{I}^2}\right)+  {\rm C_{I}^F}\ { \frac{d^2}{d\xi^2}}\ \zeta\left( -\xi,\frac{s_0^2}{\Lambda_{I}^2}\right)\,,
\eea}

\noi
where
$\zeta\left( -\xi,\frac{M_{I}^2}{\Lambda_{I}^2}\right)$ is the Hurwitz function, a generalization of the Riemann zeta function, defined by the series:
\be
\zeta(\xi,v)=\sum_{n=0}^{\infty}\frac{1}{(n+v)^{\xi}},\quad v\neq -1,-2,-3,\dots\
\ee
and its analytic continuation. For $v=1$ it reduces to the Riemann zeta function. 
The asymptotic behaviour of $\frac{1}{\pi}\Imm T^{I}(s,0)$ for $s\ra\infty$ is then governed by the inverse Mellin transform
\be\lbl{eq:invmellin}
\frac{1}{\pi}\Imm T^{I} (s,0)=\frac{1}{2\pi i} \int_{c_{\xi}-i\infty}^{c_{\xi}+i\infty} \mathrm{d}\xi \left(\frac{s}{\Lambda_{I}^2} \right)^{-\xi} 
 \left[{\rm C_{I}^V}\ \zeta\left( -\xi,\frac{M_{I}^2}{\Lambda_{I}^2}\right)+  {\rm C_{I}^F}\ { \frac{d^2}{d\xi^2}}\ \zeta\left( -\xi,\frac{s_0^2}{\Lambda_{I}^2}\right)\right]\,;
\ee
and more precisely, by 
the residues of the triple pole of the  ${\rm C_{I}^F}$ term and the single pole of the  ${\rm C_{I}^V}$ term of $\Sigma^{I}(\xi)$  at $\xi =-1$. This fixes the asymptotic behaviours of the  total $\pi^{\pm}\pi^0$ and $\pi^0 \pi^0$ cross sections in the r.h.s. of eqs.~\rf{eq:isocross1} and~\rf{eq:isocross0} as follows:

{\setl 
\bea
 \sigma_{\pi^{\pm} \pi^0}^{\rm total}(s) & \underset{{s \ra \infty}}{\thicksim}\ &  \frac{\pi }{2\Lambda_{\rho}^2}\left( {\rm C_1^F }  \log^2 \frac{s}{s_{0}}+{\rm C_1^V }\right)\,, \lbl{eq:FM1NC}\\
\sigma_{\pi^0 \pi^0}^{\rm  total}(s) & \underset{{s \ra \infty}}{\thicksim}\  & \frac{\pi }{3\Lambda_{\rho}^2}\left( {\rm C_0^F }  \log^2 \frac{s}{s_{0}}+{\rm C_0^V }\right)\,.\lbl{eq:FM0NC}
\eea}

\noi

In order to learn something about the constants $\rm C_I^{V,F}$ one can use information from the low energy behaviour of $\Ree T^{I} (s,0)$ using  a Mellin--Barnes representation of the dispersion relations in eq.~\rf{eq:F23}. This results in the expression~\cite{GdeR13} :
\be\lbl{eq:inverseMB}
\Ree T^{I}(s,0) =  
\frac{1}{2\pi i}\int_{c_{\xi}-i\infty}^{c_{\xi}+i\infty} \! \mathrm{d}\xi  \left(\frac{ s}{\Lambda_{I}^2} \right)^{2-\xi}\!\Gamma(\xi)\Gamma(1-\xi)\left[1+\frac{\pi}{\Gamma(\frac{1}{2}+\xi)\Gamma(\frac{1}{2}-\xi)} \right]\Sigma^{I} (\xi-2)\,,
\ee
where $\Sigma^{I} (\xi-2)$ is the same Mellin transform as the one defined in eq.~\rf{eq:truemellin}. The leading low energy behaviour of $\Ree T^{I}(s,0)$ is then governed by the leading singularity of the integrand in the r.h.s. of eq.~\rf{eq:inverseMB} at the left of the fundamental strip $c_{\xi}=\Ree \xi \in\ \rbrack 0,1\lbrack$, i.e. at  $\xi =0$, with the result
\be\lbl{eq:ReCs}
\Ree T^I (s,0) \underset{{s \ra 0}}{\thicksim}\ 2 \frac{s^2}{\Lambda_I ^4}\left[{\rm C_I^V}\zeta \left(2,\frac{M_I^2}{\Lambda_I^2}\right)+{\rm C_I^F}\zeta'' \left(2,\frac{s_0}{\Lambda_I^2}\right)\right]\,.
\ee
This has to match the $\chi$PT expressions in eqs.~\rf{eq:chip42} and~\rf{eq:chip43} restricted to their Large--$\Nc$ limit form i.e.,

{\setl
\bea\lbl{eq:zetaslow}
\Ree T^1 (s,0) =\Ree F_2 (s,0)+\Ree F_3(s,0) & \underset{{s \ra 0}}{\thicksim} & \frac{s^2}{f_{\pi}^4}4 l_2 (M_{\rho})\,,\\
\Ree T^0 (s,0) =\frac{3}{2}\left[3\Ree F_2 (s,0)-\Ree F_3(s,0)\right] & \underset{{s \ra 0}}{\thicksim} & \frac{s^2}{f_{\pi}^4}12\left[l_1(M_{\rho}) + l_2(M_{\rho})\right] \,.
\eea}

\noi
 This matching  gives a linear constraint between ${\rm C_I^V}$, ${\rm C_I^F}$ and the low--energy constants $l_2$ and $l_1 +l_2$:

{\setl 
\bea
32 l_2 \ \frac{M_{\rho}^4}{f_{\pi}^4} & = &  
\left[{\rm C_1^V}\zeta \left(2,\frac{M_{\rho}^2}{\Lambda_{\rho}^2}\right)+{\rm C_1^F}\zeta'' \left(2,\frac{s_0}{\Lambda_{\rho}^2}\right)\right]\,,\\
96\ (l_1 + l_2)\ \frac{M_{\rho}^4}{f_{\pi}^4} & = &  
\left[{\rm C_0^V}\zeta \left(2,\frac{M_{\sigma}^2}{\Lambda_{\sigma}^2}\right)+{\rm C_1^F}\zeta'' \left(2,\frac{s_0}{\Lambda_{\sigma}^2}\right)\right]\,.
\eea}

\noi

The  behaviour of $ \sigma_{\pi^{\pm} \pi^0}^{\rm total}(s)$ and $ \sigma_{\pi^0 \pi^0}^{\rm total}(s)$ in eqs.~\rf{eq:FM1NC} and~\rf{eq:FM0NC} at the onset of the asymptotic threshold  $s=s_0$  fixes the values of ${\rm C_1^V }$ and ${\rm C_0^V }$. We propose to identify this onset with the one of the asymptotic behaviours of $ \sigma_{\pi^{\pm} \pi^0}^{\rm total}(s)$ and $ \sigma_{\pi^0 \pi^0}^{\rm total}(s)$ in the C$\chi$QM,  evaluated  in eqs.~\rf{eq:asymppluszero} and~\rf{eq:sigma00as}. This results in the values:

{\setl
\bea
{\rm C_1^V } & = & 2 \Lambda_{\rho}^2 \frac{N_c n_f}{16\pi^2  f_{\pi}^2}
\frac{4M_Q^2}{ f_{\pi}^2}\ \frac{M_Q^2}{s_0}\ \log\frac{s_0}{M_Q^2}\,,\\
{\rm C_0^V } & = & 2 \Lambda_{\rho}^2 \frac{N_c n_f}{16\pi^2  f_{\pi}^2}
\frac{16 M_Q^2}{ f_{\pi}^2}\,.
\eea}

\noi
Given  input values for $s_0$  and $l_2$, $l_1 +l_2$ , the four  coefficients ${\rm C_1^V }$, ${\rm C_1^F }$ ${\rm C_0^V }$, ${\rm C_0^F }$ are then fixed. Obviously the onset $s_0$ has to be larger than $M_{\rho}^2$ in the $I=1$ channel and larger than $M_{\sigma}^2$ in the $I=0$. Provided one can find solutions to these constraints, the result is then a phenomenological proposal which matches the low--energy behaviour of the  the total cross sections $ \sigma_{\pi^{\pm} \pi^0}^{\rm total}(s)$ and $ \sigma_{\pi^0 \pi^0}^{\rm total}(s)$, as predicted by the C$\chi$QM, with their high--energy behaviour predicted by the Large--$\Nc$ Ansatz which we have described above. Numerical solutions to this proposal are discussed in the next section.

\vspace*{1.2cm} 
\section{\normalsize 
Numerical Results and Conclusions}
\setcounter{equation}{0}
\def\theequation{\arabic{section}.\arabic{equation}}

\noi
Several remarks concerning the issues discussed in the previous sections are in order.

\begin{itemize}
	\item {\it First, with regards to the $\chi$PT low energy constants $l_1$ and $l_2$.}
	
A recent phenomenological determination of the $l_1$ and $l_2$ constants, renormalized at the $\rho$--mass scale, gives~\cite{NPR13}:
\be\lbl{eq:phenls}
l_{1}^{\rm r}(M_{\rho})=(-5.2 \pm 0.5)\times 10^{-3}\quad\annd\quad
l_{2}^{\rm r}(M_{\rho})=(4.0\pm 1.5)\times 10^{-3}\,.
\ee
For comparison, the values predicted by the low resonance saturation of the $l_1$ and $l_2$ constants are~\cite{EGPdeR89,EGLPdeR89}:
\be\lbl{eq:LRA}
l_2 = \frac{1}{2}\frac{f_{\pi}^2}{M_{\rho}^2}\simeq 6.6\times 10^{-3}\,,\quad\annd\quad
l_1 +l_2 = \frac{1}{4}\frac{\vert c_d \vert^2}{M_{f_0}^2}\simeq 0.3\times 10^{-3}\,.
\ee
These are the values where $l_2$ is saturated by the $\rho$, with $f_{\pi}\simeq 0.088~\GeV$ i.e. the chiral limit value~\cite{Ecker13},  and $l_1 +l_2$ is saturated by the $f_0 (983)$, with $\vert c_d \vert\simeq 3.2\times 10^{-2}~\GeV$ and $M_{f_0}=0.983~\GeV$~\cite{EGPdeR89}. Within errors, these values compare rather well with the phenomenological determinations above. It is, however, far from  clear that the $f_0(983)$ is a particle one should identify with a Large--$\Nc$ ${\bar q}q$ state~\cite{PR06}. We also recall that in the C$\chi$QM (see eqs.~\rf{eq:lcqm} above) one finds:
\be\lbl{eq:l1l2QM}
l_2 = \frac{1}{3}\frac{N_c}{16\pi^2}= 6.3\times 10^{-3}\quad\annd\quad
l_1 +l_2 =\frac{1}{6}\frac{N_c}{16\pi^2}=3.2\times 10^{-3}\,.
\ee
The $l_2$ values of the C$\chi$QM  and the $\rho$ saturation approximation compare very well.	Unfortunately, the combination of low energy constants $l_1 +l_2$ determined from experiments has a large error: $l_1 +l_2 =(-1.2 \pm 1.6)\times 10^{-3}$, which makes it difficult to extract useful information on the coefficients ${\rm C_0^V}$ and ${\rm C_0^F}$ using this phenomenological input. 

\item {\it The size of $\zeta\left(2,\frac{M_I^2}{\Lambda_I^2} \right)$ and $\zeta''\left(2,\frac{s_0^2}{\Lambda_I^2} \right)$ in the r.h.s. of eq.~\rf{eq:ReCs}.}

Figure~\ref{fig:GdeRVFig2} below shows the shape of the factor  $\zeta\left(2,\frac{M_I^2}{\Lambda_I^2} \right)$ for the range  $0.2\leq \frac{M_I^2}{\Lambda_I^2}\leq 1$. The particular values we are interested in are ($\Lambda_{\rho}=\Lambda_{\sigma}=2M_{\rho}$):
\be
\zeta\left(2,\frac{M_{\rho}^2}{\Lambda_{\rho}^2}=\frac{1}{4} \right)=17.2\quad\annd\quad 
\zeta\left(2,\frac{M_{\sigma}^2}{\Lambda_{\rho}^2}=0.75 \right)=2.54\,.
\ee

Figure~\ref{fig:GdeRVFig3}  shows the shape of the factor  $\zeta''\left(2,\frac{s_0}{\Lambda_{\rho}^2} \right)$ for the range  $1\leq \frac{s_0}{M_{\rho}^2}\leq 25$. The particular values we shall be using below are:
\be
\zeta''\left(2,\frac{(2.537 M_{\rho})^2}{4 M_{\rho}^2}=1.61\right)=2.01\quad \annd\quad 
\zeta''\left(2,\frac{(3.27 M_{\sigma})^2}{4 M_{\rho}^2}=8.02\right)=1.34\,.
\ee

\begin{figure}[!h]
	\centering
	\resizebox{0.75\textwidth}{!}{\input{GdeRVFig2}}
    \caption{Shape of the factor $\zeta \left( 2, \frac{M_I^2}{\Lambda_I^2}\right)$ in the r.h.s. of eqs.~\rf{eq:ReCs}.}\label{fig:GdeRVFig2}
\end{figure}

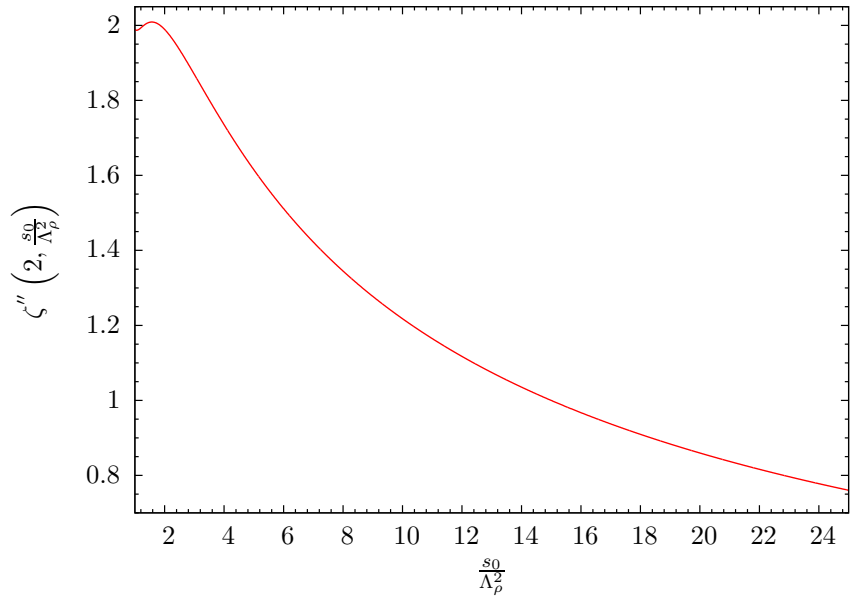
\begin{figure}[!h]
	\centering
	\resizebox{0.75\textwidth}{!}{\input{GdeRVFig3}}
    \caption{Shape of the factor $\zeta'' \left( 2, \frac{s_0}{\Lambda_\rho^2}\right)$ in the r.h.s. of eqs.~\rf{eq:ReCs}.}\label{fig:GdeRVFig3}
\end{figure}

\item {\it Fixing the constants ${\rm C_1^V}$ and ${\rm C_1^F}$ and $\sigma_{\pi^{\pm} \pi^0}^{\rm total}$.}

As a numerical example concerning the $I=1$ channel we show in figure~\ref{fig:GdeRVFig4} the total cross section $\sigma_{\pi^{\pm} \pi^0}^{\rm total}$ in millibarns versus $\sqrt{s}$ in $\GeV$ when $l_2$ is fixed to the center value of the phenomenological determination in eq.~\rf{eq:phenls} and the choice of $s_0$ is tuned to $\sqrt{s_0} =2.537 M_{\rho}$ so as to obtain a slope of the $\log^2 \frac{s}{s_0}$ term in eq.~\rf{eq:FM1NC} of the order of what is experimentally observed in the asymptotic behaviour of the total $pp$ total cross section~\cite{BH12,MS13,BH13}. The fact that this relatively simple model of Large--$\Nc$ QCD we are considering can accommodate for such a solution with a reasonable value for $s_0$ which matches the asymptotic behaviour of the C$\chi$QM is quite a remarkable fact.

\item {\it Fixing the constants ${\rm C_0^V}$ and ${\rm C_0^F}$ and $\sigma_{\pi^{\pm} \pi^0}^{\rm total}$.}

As already mentioned, the phenomenological determination of $l_1 +l_2$ has, unfortunately, a rather large error. On the other hand, the low resonance saturation value given in eq.~\rf{eq:LRA}, because of the questionable input of the scalar mass, is not reliable. The best option is to use the C$\chi$QM determination in eq.~\rf{eq:l1l2QM}, keeping in mind that this is a model--dependent input. Figure~\ref{fig:GdeRVFig5} shows the corresponding  total cross section $\sigma_{\pi^0 \pi^0}^{\rm total}$ in millibarns versus $\sqrt{s}$ in $\GeV$ for a choice of $s_0$ tuned to $\sqrt{s_0} =3.27 M_{\sigma}$ so as to obtain a slope of the $\log^2 \frac{s}{s_0}$ term in eq.~\rf{eq:FM0NC} of the order of what is experimentally observed in the asymptotic behaviour of the total $pp$ total cross section~\cite{BH12,MS13,BH13}. In spite of the fact that this is only  an illustrative example of a possible Large--$\Nc$ Ansatz, it is quite remarkable that we can find a solution which respects the constraints discussed above with rather reasonable input values.

\begin{figure}[!h]
	\centering
	\resizebox{0.75\textwidth}{!}{\input{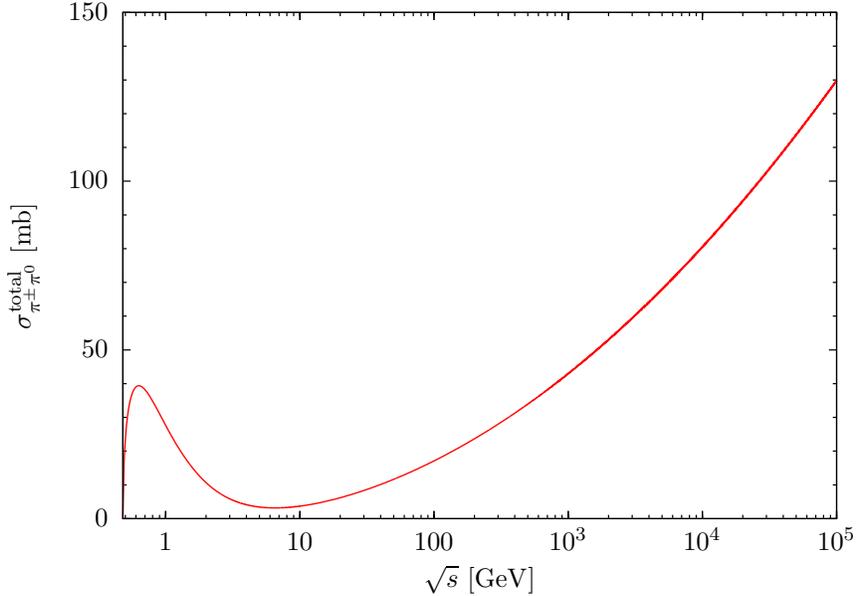}}
    \caption{$\sigma_{\pi^\pm \pi^0}^{\mathrm{total}}$ in millibarns versus $\sqrt{s}$ in $\mathrm{\GeV}$.}\label{fig:GdeRVFig4}
\end{figure}

\begin{figure}[!h]
	\centering
	\resizebox{0.75\textwidth}{!}{\input{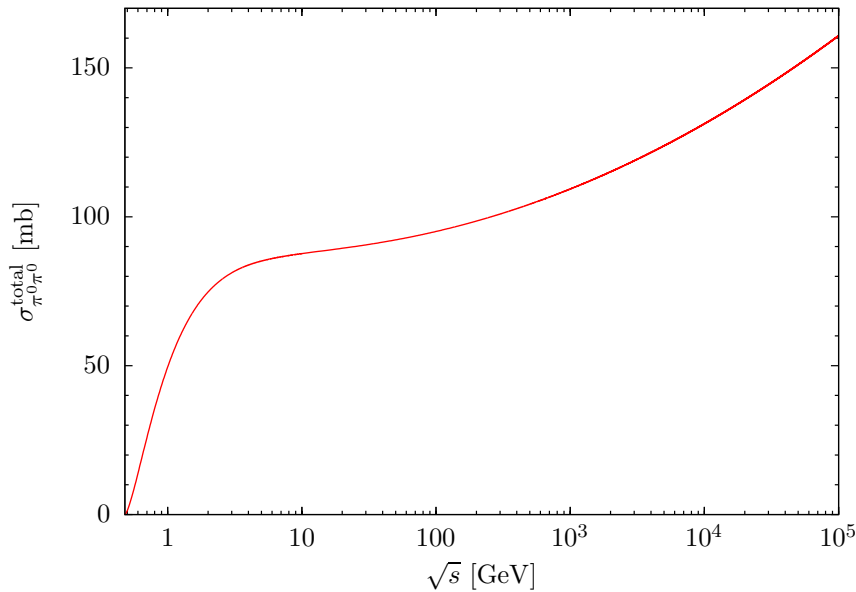}}
    \caption{$\sigma_{\pi^0 \pi^0}^{\mathrm{total}}$ in millibarns versus $\sqrt{s}$ in $\mathrm{\GeV}$.}\label{fig:GdeRVFig5}
\end{figure}

%
%
%
%
%

\end{itemize}

The main conclusion of this paper is the observation which we have discussed in Section II.1, namely the fact that -if the Froissart bound applies in QCD to $\pi\pi$ total cross sections- then the Lukaszuk-Martin coefficient $\pi / m_{\pi}^2$ of the $\log^2 s$ asymptotic behaviour cannot be the optimal one. The Lukaszuk-Martin coefficient violates both the QCD chiral behaviour and the QCD Large--$\Nc$ counting rules.

Assuming that a Froissart--like $\log^2 s$  behaviour does apply to the total $\pi\pi$ scattering cross sections in QCD, we have then shown that it is possible to construct Large--$\Nc$ Ansatz which reproduces this behaviour with coefficients which are finite  in the chiral limit and have the correct $\cO\left(1 /\Nc \right)$ counting in the Large--$\Nc$ limit. We have constructed total cross sections for $\pi^{\pm}\pi^0$ and $\pi^0\pi^0$ scattering which match the low--energy behaviour predicted by the C$\chi$QM discussed in Section III with the Large--$\Nc$ Ansatz discussed in section IV. They are shown in Figs.~\ref{fig:GdeRVFig4} and~\ref{fig:GdeRVFig5} above. The order of magnitude of these cross sections at very high energies are similar to the total $pp$ and $p{\bar p}$ scattering cross sections which are experimentally observed~\cite{BH12,MS13,BH13}. 

\vspace*{1.2cm}
 
\begin{center}
{\normalsize\bf Acknowledgments.}
\end{center}

\vspace*{0.25cm}

One of us (EdeR) is very grateful to Jos\'e Ram\'on Pelaez for clarifying issues on the phenomenological determinations of the low energy constants $l_1$ and $l_2$ as well as for guidance on the Regge literature. Two of us (DG and EdeR) are also grateful to Andr\'e Martin for correspondence and for challenging us in obtaining $\pi\pi$ cross sections in Large--$\Nc$ QCD compatible with data on $pp$ and $p{\bar p}$ total cross sections at high energies.

EdeR thanks the support of the OCEVU Labex (ANR-11-LABX-0060) and the A*MIDEX project (ANR-11-IDEX-0001-02) funded by the "Investissements d'Avenir" French government program managed by the ANR. DG's work is supported in part by the EU under Contract MTRN-CT-2006-035482 (FLAVIAnet) and by MUIR, Italy, under Project 2005-023102. GV's research  has been supported in part by the Spanish Government and ERDF funds from the EU Commission [grants FPA2007-60323, CSD2007-00042 (Consolider Project CPAN)]. GV also thanks Giancarlo D'Ambrosio and the INFN Sezione di Napoli for their hospitality and financial support. 
 


\end{document}

%% file: GdeRVFig2.tex
\begingroup%
\makeatletter%
\newcommand{\GNUPLOTspecial}{%
  \@sanitize\catcode`\%=14\relax\special}%
\setlength{\unitlength}{0.0500bp}%
\begin{picture}(7200,5040)(0,0)%
  {\GNUPLOTspecial{"
/gnudict 256 dict def
gnudict begin
%
%
/Color true def
/Blacktext true def
/Solid true def
/Dashlength 1 def
/Landscape false def
/Level1 false def
/Rounded false def
/ClipToBoundingBox false def
/SuppressPDFMark false def
/TransparentPatterns false def
/gnulinewidth 5.000 def
/userlinewidth gnulinewidth def
/Gamma 1.0 def
/BackgroundColor {-1.000 -1.000 -1.000} def
/vshift -66 def
/dl1 {
  10.0 Dashlength mul mul
  Rounded { currentlinewidth 0.75 mul sub dup 0 le { pop 0.01 } if } if
} def
/dl2 {
  10.0 Dashlength mul mul
  Rounded { currentlinewidth 0.75 mul add } if
} def
/hpt_ 31.5 def
/vpt_ 31.5 def
/hpt hpt_ def
/vpt vpt_ def
/doclip {
  ClipToBoundingBox {
    newpath 0 0 moveto 360 0 lineto 360 252 lineto 0 252 lineto closepath
    clip
  } if
} def
%
%
%
/M {moveto} bind def
/L {lineto} bind def
/R {rmoveto} bind def
/V {rlineto} bind def
/N {newpath moveto} bind def
/Z {closepath} bind def
/C {setrgbcolor} bind def
/f {rlineto fill} bind def
/g {setgray} bind def
/Gshow {show} def   
/vpt2 vpt 2 mul def
/hpt2 hpt 2 mul def
/Lshow {currentpoint stroke M 0 vshift R 
	Blacktext {gsave 0 setgray show grestore} {show} ifelse} def
/Rshow {currentpoint stroke M dup stringwidth pop neg vshift R
	Blacktext {gsave 0 setgray show grestore} {show} ifelse} def
/Cshow {currentpoint stroke M dup stringwidth pop -2 div vshift R 
	Blacktext {gsave 0 setgray show grestore} {show} ifelse} def
/UP {dup vpt_ mul /vpt exch def hpt_ mul /hpt exch def
  /hpt2 hpt 2 mul def /vpt2 vpt 2 mul def} def
/DL {Color {setrgbcolor Solid {pop []} if 0 setdash}
 {pop pop pop 0 setgray Solid {pop []} if 0 setdash} ifelse} def
/BL {stroke userlinewidth 2 mul setlinewidth
	Rounded {1 setlinejoin 1 setlinecap} if} def
/AL {stroke userlinewidth 2 div setlinewidth
	Rounded {1 setlinejoin 1 setlinecap} if} def
/UL {dup gnulinewidth mul /userlinewidth exch def
	dup 1 lt {pop 1} if 10 mul /udl exch def} def
/PL {stroke userlinewidth setlinewidth
	Rounded {1 setlinejoin 1 setlinecap} if} def
3.8 setmiterlimit
/LCw {1 1 1} def
/LCb {0 0 0} def
/LCa {0 0 0} def
/LC0 {1 0 0} def
/LC1 {0 1 0} def
/LC2 {0 0 1} def
/LC3 {1 0 1} def
/LC4 {0 1 1} def
/LC5 {1 1 0} def
/LC6 {0 0 0} def
/LC7 {1 0.3 0} def
/LC8 {0.5 0.5 0.5} def
/LTw {PL [] 1 setgray} def
/LTb {BL [] LCb DL} def
/LTa {AL [1 udl mul 2 udl mul] 0 setdash LCa setrgbcolor} def
/LT0 {PL [] LC0 DL} def
/LT1 {PL [4 dl1 2 dl2] LC1 DL} def
/LT2 {PL [2 dl1 3 dl2] LC2 DL} def
/LT3 {PL [1 dl1 1.5 dl2] LC3 DL} def
/LT4 {PL [6 dl1 2 dl2 1 dl1 2 dl2] LC4 DL} def
/LT5 {PL [3 dl1 3 dl2 1 dl1 3 dl2] LC5 DL} def
/LT6 {PL [2 dl1 2 dl2 2 dl1 6 dl2] LC6 DL} def
/LT7 {PL [1 dl1 2 dl2 6 dl1 2 dl2 1 dl1 2 dl2] LC7 DL} def
/LT8 {PL [2 dl1 2 dl2 2 dl1 2 dl2 2 dl1 2 dl2 2 dl1 4 dl2] LC8 DL} def
/Pnt {stroke [] 0 setdash gsave 1 setlinecap M 0 0 V stroke grestore} def
/Dia {stroke [] 0 setdash 2 copy vpt add M
  hpt neg vpt neg V hpt vpt neg V
  hpt vpt V hpt neg vpt V closepath stroke
  Pnt} def
/Pls {stroke [] 0 setdash vpt sub M 0 vpt2 V
  currentpoint stroke M
  hpt neg vpt neg R hpt2 0 V stroke
 } def
/Box {stroke [] 0 setdash 2 copy exch hpt sub exch vpt add M
  0 vpt2 neg V hpt2 0 V 0 vpt2 V
  hpt2 neg 0 V closepath stroke
  Pnt} def
/Crs {stroke [] 0 setdash exch hpt sub exch vpt add M
  hpt2 vpt2 neg V currentpoint stroke M
  hpt2 neg 0 R hpt2 vpt2 V stroke} def
/TriU {stroke [] 0 setdash 2 copy vpt 1.12 mul add M
  hpt neg vpt -1.62 mul V
  hpt 2 mul 0 V
  hpt neg vpt 1.62 mul V closepath stroke
  Pnt} def
/Star {2 copy Pls Crs} def
/BoxF {stroke [] 0 setdash exch hpt sub exch vpt add M
  0 vpt2 neg V hpt2 0 V 0 vpt2 V
  hpt2 neg 0 V closepath fill} def
/TriUF {stroke [] 0 setdash vpt 1.12 mul add M
  hpt neg vpt -1.62 mul V
  hpt 2 mul 0 V
  hpt neg vpt 1.62 mul V closepath fill} def
/TriD {stroke [] 0 setdash 2 copy vpt 1.12 mul sub M
  hpt neg vpt 1.62 mul V
  hpt 2 mul 0 V
  hpt neg vpt -1.62 mul V closepath stroke
  Pnt} def
/TriDF {stroke [] 0 setdash vpt 1.12 mul sub M
  hpt neg vpt 1.62 mul V
  hpt 2 mul 0 V
  hpt neg vpt -1.62 mul V closepath fill} def
/DiaF {stroke [] 0 setdash vpt add M
  hpt neg vpt neg V hpt vpt neg V
  hpt vpt V hpt neg vpt V closepath fill} def
/Pent {stroke [] 0 setdash 2 copy gsave
  translate 0 hpt M 4 {72 rotate 0 hpt L} repeat
  closepath stroke grestore Pnt} def
/PentF {stroke [] 0 setdash gsave
  translate 0 hpt M 4 {72 rotate 0 hpt L} repeat
  closepath fill grestore} def
/Circle {stroke [] 0 setdash 2 copy
  hpt 0 360 arc stroke Pnt} def
/CircleF {stroke [] 0 setdash hpt 0 360 arc fill} def
/C0 {BL [] 0 setdash 2 copy moveto vpt 90 450 arc} bind def
/C1 {BL [] 0 setdash 2 copy moveto
	2 copy vpt 0 90 arc closepath fill
	vpt 0 360 arc closepath} bind def
/C2 {BL [] 0 setdash 2 copy moveto
	2 copy vpt 90 180 arc closepath fill
	vpt 0 360 arc closepath} bind def
/C3 {BL [] 0 setdash 2 copy moveto
	2 copy vpt 0 180 arc closepath fill
	vpt 0 360 arc closepath} bind def
/C4 {BL [] 0 setdash 2 copy moveto
	2 copy vpt 180 270 arc closepath fill
	vpt 0 360 arc closepath} bind def
/C5 {BL [] 0 setdash 2 copy moveto
	2 copy vpt 0 90 arc
	2 copy moveto
	2 copy vpt 180 270 arc closepath fill
	vpt 0 360 arc} bind def
/C6 {BL [] 0 setdash 2 copy moveto
	2 copy vpt 90 270 arc closepath fill
	vpt 0 360 arc closepath} bind def
/C7 {BL [] 0 setdash 2 copy moveto
	2 copy vpt 0 270 arc closepath fill
	vpt 0 360 arc closepath} bind def
/C8 {BL [] 0 setdash 2 copy moveto
	2 copy vpt 270 360 arc closepath fill
	vpt 0 360 arc closepath} bind def
/C9 {BL [] 0 setdash 2 copy moveto
	2 copy vpt 270 450 arc closepath fill
	vpt 0 360 arc closepath} bind def
/C10 {BL [] 0 setdash 2 copy 2 copy moveto vpt 270 360 arc closepath fill
	2 copy moveto
	2 copy vpt 90 180 arc closepath fill
	vpt 0 360 arc closepath} bind def
/C11 {BL [] 0 setdash 2 copy moveto
	2 copy vpt 0 180 arc closepath fill
	2 copy moveto
	2 copy vpt 270 360 arc closepath fill
	vpt 0 360 arc closepath} bind def
/C12 {BL [] 0 setdash 2 copy moveto
	2 copy vpt 180 360 arc closepath fill
	vpt 0 360 arc closepath} bind def
/C13 {BL [] 0 setdash 2 copy moveto
	2 copy vpt 0 90 arc closepath fill
	2 copy moveto
	2 copy vpt 180 360 arc closepath fill
	vpt 0 360 arc closepath} bind def
/C14 {BL [] 0 setdash 2 copy moveto
	2 copy vpt 90 360 arc closepath fill
	vpt 0 360 arc} bind def
/C15 {BL [] 0 setdash 2 copy vpt 0 360 arc closepath fill
	vpt 0 360 arc closepath} bind def
/Rec {newpath 4 2 roll moveto 1 index 0 rlineto 0 exch rlineto
	neg 0 rlineto closepath} bind def
/Square {dup Rec} bind def
/Bsquare {vpt sub exch vpt sub exch vpt2 Square} bind def
/S0 {BL [] 0 setdash 2 copy moveto 0 vpt rlineto BL Bsquare} bind def
/S1 {BL [] 0 setdash 2 copy vpt Square fill Bsquare} bind def
/S2 {BL [] 0 setdash 2 copy exch vpt sub exch vpt Square fill Bsquare} bind def
/S3 {BL [] 0 setdash 2 copy exch vpt sub exch vpt2 vpt Rec fill Bsquare} bind def
/S4 {BL [] 0 setdash 2 copy exch vpt sub exch vpt sub vpt Square fill Bsquare} bind def
/S5 {BL [] 0 setdash 2 copy 2 copy vpt Square fill
	exch vpt sub exch vpt sub vpt Square fill Bsquare} bind def
/S6 {BL [] 0 setdash 2 copy exch vpt sub exch vpt sub vpt vpt2 Rec fill Bsquare} bind def
/S7 {BL [] 0 setdash 2 copy exch vpt sub exch vpt sub vpt vpt2 Rec fill
	2 copy vpt Square fill Bsquare} bind def
/S8 {BL [] 0 setdash 2 copy vpt sub vpt Square fill Bsquare} bind def
/S9 {BL [] 0 setdash 2 copy vpt sub vpt vpt2 Rec fill Bsquare} bind def
/S10 {BL [] 0 setdash 2 copy vpt sub vpt Square fill 2 copy exch vpt sub exch vpt Square fill
	Bsquare} bind def
/S11 {BL [] 0 setdash 2 copy vpt sub vpt Square fill 2 copy exch vpt sub exch vpt2 vpt Rec fill
	Bsquare} bind def
/S12 {BL [] 0 setdash 2 copy exch vpt sub exch vpt sub vpt2 vpt Rec fill Bsquare} bind def
/S13 {BL [] 0 setdash 2 copy exch vpt sub exch vpt sub vpt2 vpt Rec fill
	2 copy vpt Square fill Bsquare} bind def
/S14 {BL [] 0 setdash 2 copy exch vpt sub exch vpt sub vpt2 vpt Rec fill
	2 copy exch vpt sub exch vpt Square fill Bsquare} bind def
/S15 {BL [] 0 setdash 2 copy Bsquare fill Bsquare} bind def
/D0 {gsave translate 45 rotate 0 0 S0 stroke grestore} bind def
/D1 {gsave translate 45 rotate 0 0 S1 stroke grestore} bind def
/D2 {gsave translate 45 rotate 0 0 S2 stroke grestore} bind def
/D3 {gsave translate 45 rotate 0 0 S3 stroke grestore} bind def
/D4 {gsave translate 45 rotate 0 0 S4 stroke grestore} bind def
/D5 {gsave translate 45 rotate 0 0 S5 stroke grestore} bind def
/D6 {gsave translate 45 rotate 0 0 S6 stroke grestore} bind def
/D7 {gsave translate 45 rotate 0 0 S7 stroke grestore} bind def
/D8 {gsave translate 45 rotate 0 0 S8 stroke grestore} bind def
/D9 {gsave translate 45 rotate 0 0 S9 stroke grestore} bind def
/D10 {gsave translate 45 rotate 0 0 S10 stroke grestore} bind def
/D11 {gsave translate 45 rotate 0 0 S11 stroke grestore} bind def
/D12 {gsave translate 45 rotate 0 0 S12 stroke grestore} bind def
/D13 {gsave translate 45 rotate 0 0 S13 stroke grestore} bind def
/D14 {gsave translate 45 rotate 0 0 S14 stroke grestore} bind def
/D15 {gsave translate 45 rotate 0 0 S15 stroke grestore} bind def
/DiaE {stroke [] 0 setdash vpt add M
  hpt neg vpt neg V hpt vpt neg V
  hpt vpt V hpt neg vpt V closepath stroke} def
/BoxE {stroke [] 0 setdash exch hpt sub exch vpt add M
  0 vpt2 neg V hpt2 0 V 0 vpt2 V
  hpt2 neg 0 V closepath stroke} def
/TriUE {stroke [] 0 setdash vpt 1.12 mul add M
  hpt neg vpt -1.62 mul V
  hpt 2 mul 0 V
  hpt neg vpt 1.62 mul V closepath stroke} def
/TriDE {stroke [] 0 setdash vpt 1.12 mul sub M
  hpt neg vpt 1.62 mul V
  hpt 2 mul 0 V
  hpt neg vpt -1.62 mul V closepath stroke} def
/PentE {stroke [] 0 setdash gsave
  translate 0 hpt M 4 {72 rotate 0 hpt L} repeat
  closepath stroke grestore} def
/CircE {stroke [] 0 setdash 
  hpt 0 360 arc stroke} def
/Opaque {gsave closepath 1 setgray fill grestore 0 setgray closepath} def
/DiaW {stroke [] 0 setdash vpt add M
  hpt neg vpt neg V hpt vpt neg V
  hpt vpt V hpt neg vpt V Opaque stroke} def
/BoxW {stroke [] 0 setdash exch hpt sub exch vpt add M
  0 vpt2 neg V hpt2 0 V 0 vpt2 V
  hpt2 neg 0 V Opaque stroke} def
/TriUW {stroke [] 0 setdash vpt 1.12 mul add M
  hpt neg vpt -1.62 mul V
  hpt 2 mul 0 V
  hpt neg vpt 1.62 mul V Opaque stroke} def
/TriDW {stroke [] 0 setdash vpt 1.12 mul sub M
  hpt neg vpt 1.62 mul V
  hpt 2 mul 0 V
  hpt neg vpt -1.62 mul V Opaque stroke} def
/PentW {stroke [] 0 setdash gsave
  translate 0 hpt M 4 {72 rotate 0 hpt L} repeat
  Opaque stroke grestore} def
/CircW {stroke [] 0 setdash 
  hpt 0 360 arc Opaque stroke} def
/BoxFill {gsave Rec 1 setgray fill grestore} def
/Density {
  /Fillden exch def
  currentrgbcolor
  /ColB exch def /ColG exch def /ColR exch def
  /ColR ColR Fillden mul Fillden sub 1 add def
  /ColG ColG Fillden mul Fillden sub 1 add def
  /ColB ColB Fillden mul Fillden sub 1 add def
  ColR ColG ColB setrgbcolor} def
/BoxColFill {gsave Rec PolyFill} def
/PolyFill {gsave Density fill grestore grestore} def
/h {rlineto rlineto rlineto gsave closepath fill grestore} bind def
%
%
/PatternFill {gsave /PFa [ 9 2 roll ] def
  PFa 0 get PFa 2 get 2 div add PFa 1 get PFa 3 get 2 div add translate
  PFa 2 get -2 div PFa 3 get -2 div PFa 2 get PFa 3 get Rec
  TransparentPatterns {} {gsave 1 setgray fill grestore} ifelse
  clip
  currentlinewidth 0.5 mul setlinewidth
  /PFs PFa 2 get dup mul PFa 3 get dup mul add sqrt def
  0 0 M PFa 5 get rotate PFs -2 div dup translate
  0 1 PFs PFa 4 get div 1 add floor cvi
	{PFa 4 get mul 0 M 0 PFs V} for
  0 PFa 6 get ne {
	0 1 PFs PFa 4 get div 1 add floor cvi
	{PFa 4 get mul 0 2 1 roll M PFs 0 V} for
 } if
  stroke grestore} def
/languagelevel where
 {pop languagelevel} {1} ifelse
 2 lt
	{/InterpretLevel1 true def}
	{/InterpretLevel1 Level1 def}
 ifelse
%
%
/Level2PatternFill {
/Tile8x8 {/PaintType 2 /PatternType 1 /TilingType 1 /BBox [0 0 8 8] /XStep 8 /YStep 8}
	bind def
/KeepColor {currentrgbcolor [/Pattern /DeviceRGB] setcolorspace} bind def
<< Tile8x8
 /PaintProc {0.5 setlinewidth pop 0 0 M 8 8 L 0 8 M 8 0 L stroke} 
>> matrix makepattern
/Pat1 exch def
<< Tile8x8
 /PaintProc {0.5 setlinewidth pop 0 0 M 8 8 L 0 8 M 8 0 L stroke
	0 4 M 4 8 L 8 4 L 4 0 L 0 4 L stroke}
>> matrix makepattern
/Pat2 exch def
<< Tile8x8
 /PaintProc {0.5 setlinewidth pop 0 0 M 0 8 L
	8 8 L 8 0 L 0 0 L fill}
>> matrix makepattern
/Pat3 exch def
<< Tile8x8
 /PaintProc {0.5 setlinewidth pop -4 8 M 8 -4 L
	0 12 M 12 0 L stroke}
>> matrix makepattern
/Pat4 exch def
<< Tile8x8
 /PaintProc {0.5 setlinewidth pop -4 0 M 8 12 L
	0 -4 M 12 8 L stroke}
>> matrix makepattern
/Pat5 exch def
<< Tile8x8
 /PaintProc {0.5 setlinewidth pop -2 8 M 4 -4 L
	0 12 M 8 -4 L 4 12 M 10 0 L stroke}
>> matrix makepattern
/Pat6 exch def
<< Tile8x8
 /PaintProc {0.5 setlinewidth pop -2 0 M 4 12 L
	0 -4 M 8 12 L 4 -4 M 10 8 L stroke}
>> matrix makepattern
/Pat7 exch def
<< Tile8x8
 /PaintProc {0.5 setlinewidth pop 8 -2 M -4 4 L
	12 0 M -4 8 L 12 4 M 0 10 L stroke}
>> matrix makepattern
/Pat8 exch def
<< Tile8x8
 /PaintProc {0.5 setlinewidth pop 0 -2 M 12 4 L
	-4 0 M 12 8 L -4 4 M 8 10 L stroke}
>> matrix makepattern
/Pat9 exch def
/Pattern1 {PatternBgnd KeepColor Pat1 setpattern} bind def
/Pattern2 {PatternBgnd KeepColor Pat2 setpattern} bind def
/Pattern3 {PatternBgnd KeepColor Pat3 setpattern} bind def
/Pattern4 {PatternBgnd KeepColor Landscape {Pat5} {Pat4} ifelse setpattern} bind def
/Pattern5 {PatternBgnd KeepColor Landscape {Pat4} {Pat5} ifelse setpattern} bind def
/Pattern6 {PatternBgnd KeepColor Landscape {Pat9} {Pat6} ifelse setpattern} bind def
/Pattern7 {PatternBgnd KeepColor Landscape {Pat8} {Pat7} ifelse setpattern} bind def
} def
%
%
%
/PatternBgnd {
  TransparentPatterns {} {gsave 1 setgray fill grestore} ifelse
} def
%
%
/Level1PatternFill {
/Pattern1 {0.250 Density} bind def
/Pattern2 {0.500 Density} bind def
/Pattern3 {0.750 Density} bind def
/Pattern4 {0.125 Density} bind def
/Pattern5 {0.375 Density} bind def
/Pattern6 {0.625 Density} bind def
/Pattern7 {0.875 Density} bind def
} def
%
%
Level1 {Level1PatternFill} {Level2PatternFill} ifelse
/Symbol-Oblique /Symbol findfont [1 0 .167 1 0 0] makefont
dup length dict begin {1 index /FID eq {pop pop} {def} ifelse} forall
currentdict end definefont pop
Level1 SuppressPDFMark or 
{} {
/SDict 10 dict def
systemdict /pdfmark known not {
  userdict /pdfmark systemdict /cleartomark get put
} if
SDict begin [
  /Title (fig2.tex)
  /Subject (gnuplot plot)
  /Creator (gnuplot 4.6 patchlevel 4)
  /Author (gregoryvulvert)
  /CreationDate (Tue Dec  3 11:39:07 2013)
  /DOCINFO pdfmark
end
} ifelse
end
gnudict begin
gsave
doclip
0 0 translate
0.050 0.050 scale
0 setgray
newpath
BackgroundColor 0 lt 3 1 roll 0 lt exch 0 lt or or not {BackgroundColor C 1.000 0 0 7200.00 5040.00 BoxColFill} if
1.000 UL
LTb
LCb setrgbcolor
860 640 M
63 0 V
5916 0 R
-63 0 V
stroke
LTb
LCb setrgbcolor
860 779 M
31 0 V
5948 0 R
-31 0 V
860 917 M
31 0 V
5948 0 R
-31 0 V
860 1056 M
31 0 V
5948 0 R
-31 0 V
860 1195 M
31 0 V
5948 0 R
-31 0 V
860 1333 M
63 0 V
5916 0 R
-63 0 V
stroke
LTb
LCb setrgbcolor
860 1472 M
31 0 V
5948 0 R
-31 0 V
860 1610 M
31 0 V
5948 0 R
-31 0 V
860 1749 M
31 0 V
5948 0 R
-31 0 V
860 1888 M
31 0 V
5948 0 R
-31 0 V
860 2026 M
63 0 V
5916 0 R
-63 0 V
stroke
LTb
LCb setrgbcolor
860 2165 M
31 0 V
5948 0 R
-31 0 V
860 2304 M
31 0 V
5948 0 R
-31 0 V
860 2442 M
31 0 V
5948 0 R
-31 0 V
860 2581 M
31 0 V
5948 0 R
-31 0 V
860 2720 M
63 0 V
5916 0 R
-63 0 V
stroke
LTb
LCb setrgbcolor
860 2858 M
31 0 V
5948 0 R
-31 0 V
860 2997 M
31 0 V
5948 0 R
-31 0 V
860 3135 M
31 0 V
5948 0 R
-31 0 V
860 3274 M
31 0 V
5948 0 R
-31 0 V
860 3413 M
63 0 V
5916 0 R
-63 0 V
stroke
LTb
LCb setrgbcolor
860 3551 M
31 0 V
5948 0 R
-31 0 V
860 3690 M
31 0 V
5948 0 R
-31 0 V
860 3829 M
31 0 V
5948 0 R
-31 0 V
860 3967 M
31 0 V
5948 0 R
-31 0 V
860 4106 M
63 0 V
5916 0 R
-63 0 V
stroke
LTb
LCb setrgbcolor
860 4244 M
31 0 V
5948 0 R
-31 0 V
860 4383 M
31 0 V
5948 0 R
-31 0 V
860 4522 M
31 0 V
5948 0 R
-31 0 V
860 4660 M
31 0 V
5948 0 R
-31 0 V
860 4799 M
63 0 V
5916 0 R
-63 0 V
stroke
LTb
LCb setrgbcolor
860 640 M
0 63 V
0 4096 R
0 -63 V
stroke
LTb
LCb setrgbcolor
1009 640 M
0 31 V
0 4128 R
0 -31 V
1159 640 M
0 31 V
0 4128 R
0 -31 V
1308 640 M
0 31 V
0 4128 R
0 -31 V
1458 640 M
0 31 V
0 4128 R
0 -31 V
1607 640 M
0 63 V
0 4096 R
0 -63 V
stroke
LTb
LCb setrgbcolor
1757 640 M
0 31 V
0 4128 R
0 -31 V
1906 640 M
0 31 V
0 4128 R
0 -31 V
2056 640 M
0 31 V
0 4128 R
0 -31 V
2205 640 M
0 31 V
0 4128 R
0 -31 V
2355 640 M
0 63 V
0 4096 R
0 -63 V
stroke
LTb
LCb setrgbcolor
2504 640 M
0 31 V
0 4128 R
0 -31 V
2654 640 M
0 31 V
0 4128 R
0 -31 V
2803 640 M
0 31 V
0 4128 R
0 -31 V
2953 640 M
0 31 V
0 4128 R
0 -31 V
3102 640 M
0 63 V
0 4096 R
0 -63 V
stroke
LTb
LCb setrgbcolor
3252 640 M
0 31 V
0 4128 R
0 -31 V
3401 640 M
0 31 V
0 4128 R
0 -31 V
3551 640 M
0 31 V
0 4128 R
0 -31 V
3700 640 M
0 31 V
0 4128 R
0 -31 V
3849 640 M
0 63 V
0 4096 R
0 -63 V
stroke
LTb
LCb setrgbcolor
3999 640 M
0 31 V
0 4128 R
0 -31 V
4148 640 M
0 31 V
0 4128 R
0 -31 V
4298 640 M
0 31 V
0 4128 R
0 -31 V
4447 640 M
0 31 V
0 4128 R
0 -31 V
4597 640 M
0 63 V
0 4096 R
0 -63 V
stroke
LTb
LCb setrgbcolor
4746 640 M
0 31 V
0 4128 R
0 -31 V
4896 640 M
0 31 V
0 4128 R
0 -31 V
5045 640 M
0 31 V
0 4128 R
0 -31 V
5195 640 M
0 31 V
0 4128 R
0 -31 V
5344 640 M
0 63 V
0 4096 R
0 -63 V
stroke
LTb
LCb setrgbcolor
5494 640 M
0 31 V
0 4128 R
0 -31 V
5643 640 M
0 31 V
0 4128 R
0 -31 V
5793 640 M
0 31 V
0 4128 R
0 -31 V
5942 640 M
0 31 V
0 4128 R
0 -31 V
6092 640 M
0 63 V
0 4096 R
0 -63 V
stroke
LTb
LCb setrgbcolor
6241 640 M
0 31 V
0 4128 R
0 -31 V
6391 640 M
0 31 V
0 4128 R
0 -31 V
6540 640 M
0 31 V
0 4128 R
0 -31 V
6690 640 M
0 31 V
0 4128 R
0 -31 V
6839 640 M
0 63 V
0 4096 R
0 -63 V
stroke
LTb
LCb setrgbcolor
1.000 UL
LTb
LCb setrgbcolor
860 4799 N
860 640 L
5979 0 V
0 4159 V
-5979 0 V
Z stroke
LCb setrgbcolor
LTb
LCb setrgbcolor
LTb
1.000 UP
1.000 UL
LTb
LCb setrgbcolor
2.200 UL
LT0
1.00 0.00 0.00 C 860 4282 M
7 -35 V
8 -34 V
7 -34 V
8 -33 V
7 -32 V
8 -33 V
7 -31 V
8 -32 V
7 -30 V
8 -31 V
7 -30 V
8 -29 V
7 -29 V
8 -29 V
7 -28 V
8 -28 V
7 -28 V
8 -27 V
7 -27 V
7 -26 V
8 -26 V
7 -26 V
8 -25 V
7 -25 V
8 -25 V
7 -24 V
8 -24 V
7 -24 V
8 -23 V
7 -24 V
8 -22 V
7 -23 V
8 -22 V
7 -22 V
8 -22 V
7 -21 V
8 -21 V
7 -21 V
7 -21 V
8 -20 V
7 -20 V
8 -20 V
7 -20 V
8 -19 V
7 -19 V
8 -19 V
7 -19 V
8 -19 V
7 -18 V
8 -18 V
7 -18 V
8 -17 V
7 -18 V
8 -17 V
7 -17 V
8 -17 V
7 -16 V
7 -17 V
8 -16 V
7 -16 V
8 -16 V
7 -16 V
8 -15 V
7 -15 V
8 -16 V
7 -15 V
8 -14 V
7 -15 V
8 -15 V
7 -14 V
8 -14 V
7 -14 V
8 -14 V
7 -14 V
8 -13 V
7 -14 V
7 -13 V
8 -13 V
7 -13 V
8 -13 V
7 -13 V
8 -12 V
7 -13 V
8 -12 V
7 -12 V
8 -13 V
7 -11 V
8 -12 V
7 -12 V
8 -12 V
7 -11 V
8 -11 V
7 -12 V
8 -11 V
7 -11 V
7 -11 V
8 -11 V
7 -10 V
8 -11 V
7 -10 V
8 -11 V
7 -10 V
8 -10 V
7 -10 V
stroke 1637 2297 M
8 -10 V
7 -10 V
8 -10 V
7 -10 V
8 -9 V
7 -10 V
8 -9 V
7 -10 V
8 -9 V
7 -9 V
7 -9 V
8 -9 V
7 -9 V
8 -9 V
7 -9 V
8 -8 V
7 -9 V
8 -8 V
7 -9 V
8 -8 V
7 -8 V
8 -8 V
7 -9 V
8 -8 V
7 -8 V
8 -8 V
7 -7 V
8 -8 V
7 -8 V
7 -7 V
8 -8 V
7 -8 V
8 -7 V
7 -7 V
8 -8 V
7 -7 V
8 -7 V
7 -7 V
8 -7 V
7 -7 V
8 -7 V
7 -7 V
8 -7 V
7 -7 V
8 -6 V
7 -7 V
8 -7 V
7 -6 V
7 -7 V
8 -6 V
7 -6 V
8 -7 V
7 -6 V
8 -6 V
7 -6 V
8 -6 V
7 -7 V
8 -6 V
7 -6 V
8 -5 V
7 -6 V
8 -6 V
7 -6 V
8 -6 V
7 -5 V
8 -6 V
7 -6 V
7 -5 V
8 -6 V
7 -5 V
8 -6 V
7 -5 V
8 -5 V
7 -6 V
8 -5 V
7 -5 V
8 -5 V
7 -5 V
8 -5 V
7 -5 V
8 -5 V
7 -5 V
8 -5 V
7 -5 V
8 -5 V
7 -5 V
7 -5 V
8 -5 V
7 -4 V
8 -5 V
7 -5 V
8 -4 V
7 -5 V
8 -4 V
7 -5 V
8 -4 V
7 -5 V
8 -4 V
7 -5 V
8 -4 V
7 -4 V
8 -5 V
7 -4 V
8 -4 V
stroke 2415 1614 M
7 -4 V
7 -5 V
8 -4 V
7 -4 V
8 -4 V
7 -4 V
8 -4 V
7 -4 V
8 -4 V
7 -4 V
8 -4 V
7 -4 V
8 -4 V
7 -3 V
8 -4 V
7 -4 V
8 -4 V
7 -3 V
8 -4 V
7 -4 V
7 -3 V
8 -4 V
7 -4 V
8 -3 V
7 -4 V
8 -3 V
7 -4 V
8 -3 V
7 -4 V
8 -3 V
7 -4 V
8 -3 V
7 -3 V
8 -4 V
7 -3 V
8 -3 V
7 -4 V
8 -3 V
7 -3 V
7 -3 V
8 -4 V
7 -3 V
8 -3 V
7 -3 V
8 -3 V
7 -3 V
8 -3 V
7 -3 V
8 -3 V
7 -3 V
8 -3 V
7 -3 V
8 -3 V
7 -3 V
8 -3 V
7 -3 V
8 -3 V
7 -3 V
7 -3 V
8 -3 V
7 -2 V
8 -3 V
7 -3 V
8 -3 V
7 -3 V
8 -2 V
7 -3 V
8 -3 V
7 -2 V
8 -3 V
7 -3 V
8 -2 V
7 -3 V
8 -3 V
7 -2 V
8 -3 V
7 -2 V
7 -3 V
8 -2 V
7 -3 V
8 -2 V
7 -3 V
8 -2 V
7 -3 V
8 -2 V
7 -3 V
8 -2 V
7 -2 V
8 -3 V
7 -2 V
8 -3 V
7 -2 V
8 -2 V
7 -3 V
8 -2 V
7 -2 V
7 -2 V
8 -3 V
7 -2 V
8 -2 V
7 -2 V
8 -3 V
7 -2 V
8 -2 V
stroke 3192 1297 M
7 -2 V
8 -2 V
7 -2 V
8 -3 V
7 -2 V
8 -2 V
7 -2 V
8 -2 V
7 -2 V
8 -2 V
7 -2 V
7 -2 V
8 -2 V
7 -2 V
8 -2 V
7 -2 V
8 -2 V
7 -2 V
8 -2 V
7 -2 V
8 -2 V
7 -2 V
8 -2 V
7 -2 V
8 -2 V
7 -2 V
8 -2 V
7 -1 V
8 -2 V
7 -2 V
7 -2 V
8 -2 V
7 -2 V
8 -1 V
7 -2 V
8 -2 V
7 -2 V
8 -2 V
7 -1 V
8 -2 V
7 -2 V
8 -2 V
7 -1 V
8 -2 V
7 -2 V
8 -2 V
7 -1 V
8 -2 V
7 -2 V
7 -1 V
8 -2 V
7 -2 V
8 -1 V
7 -2 V
8 -1 V
7 -2 V
8 -2 V
7 -1 V
8 -2 V
7 -1 V
8 -2 V
7 -2 V
8 -1 V
7 -2 V
8 -1 V
7 -2 V
8 -1 V
7 -2 V
7 -1 V
8 -2 V
7 -1 V
8 -2 V
7 -1 V
8 -2 V
7 -1 V
8 -2 V
7 -1 V
8 -2 V
7 -1 V
8 -2 V
7 -1 V
8 -1 V
7 -2 V
8 -1 V
7 -2 V
8 -1 V
7 -2 V
8 -1 V
7 -1 V
7 -2 V
8 -1 V
7 -1 V
8 -2 V
7 -1 V
8 -1 V
7 -2 V
8 -1 V
7 -1 V
8 -2 V
7 -1 V
8 -1 V
7 -2 V
8 -1 V
7 -1 V
stroke 3969 1123 M
8 -1 V
7 -2 V
8 -1 V
7 -1 V
7 -2 V
8 -1 V
7 -1 V
8 -1 V
7 -1 V
8 -2 V
7 -1 V
8 -1 V
7 -1 V
8 -2 V
7 -1 V
8 -1 V
7 -1 V
8 -1 V
7 -2 V
8 -1 V
7 -1 V
8 -1 V
7 -1 V
7 -1 V
8 -1 V
7 -2 V
8 -1 V
7 -1 V
8 -1 V
7 -1 V
8 -1 V
7 -1 V
8 -1 V
7 -2 V
8 -1 V
7 -1 V
8 -1 V
7 -1 V
8 -1 V
7 -1 V
8 -1 V
7 -1 V
7 -1 V
8 -1 V
7 -1 V
8 -1 V
7 -1 V
8 -1 V
7 -2 V
8 -1 V
7 -1 V
8 -1 V
7 -1 V
8 -1 V
7 -1 V
8 -1 V
7 -1 V
8 -1 V
7 -1 V
8 -1 V
7 -1 V
7 -1 V
8 -1 V
7 -1 V
8 0 V
7 -1 V
8 -1 V
7 -1 V
8 -1 V
7 -1 V
8 -1 V
7 -1 V
8 -1 V
7 -1 V
8 -1 V
7 -1 V
8 -1 V
7 -1 V
8 -1 V
7 -1 V
7 0 V
8 -1 V
7 -1 V
8 -1 V
7 -1 V
8 -1 V
7 -1 V
8 -1 V
7 -1 V
8 0 V
7 -1 V
8 -1 V
7 -1 V
8 -1 V
7 -1 V
8 -1 V
7 0 V
8 -1 V
7 -1 V
7 -1 V
8 -1 V
7 -1 V
8 0 V
7 -1 V
stroke 4746 1016 M
8 -1 V
7 -1 V
8 -1 V
7 -1 V
8 0 V
7 -1 V
8 -1 V
7 -1 V
8 -1 V
7 0 V
8 -1 V
7 -1 V
8 -1 V
7 0 V
7 -1 V
8 -1 V
7 -1 V
8 -1 V
7 0 V
8 -1 V
7 -1 V
8 -1 V
7 0 V
8 -1 V
7 -1 V
8 -1 V
7 0 V
8 -1 V
7 -1 V
8 -1 V
7 0 V
8 -1 V
7 -1 V
7 -1 V
8 0 V
7 -1 V
8 -1 V
7 0 V
8 -1 V
7 -1 V
8 -1 V
7 0 V
8 -1 V
7 -1 V
8 0 V
7 -1 V
8 -1 V
7 0 V
8 -1 V
7 -1 V
8 0 V
7 -1 V
7 -1 V
8 0 V
7 -1 V
8 -1 V
7 -1 V
8 0 V
7 -1 V
8 0 V
7 -1 V
8 -1 V
7 0 V
8 -1 V
7 -1 V
8 0 V
7 -1 V
8 -1 V
7 0 V
8 -1 V
7 -1 V
7 0 V
8 -1 V
7 0 V
8 -1 V
7 -1 V
8 0 V
7 -1 V
8 -1 V
7 0 V
8 -1 V
7 0 V
8 -1 V
7 -1 V
8 0 V
7 -1 V
8 0 V
7 -1 V
8 -1 V
7 0 V
7 -1 V
8 0 V
7 -1 V
8 -1 V
7 0 V
8 -1 V
7 0 V
8 -1 V
7 -1 V
8 0 V
7 -1 V
8 0 V
7 -1 V
8 0 V
stroke 5524 945 M
7 -1 V
8 0 V
7 -1 V
8 -1 V
7 0 V
7 -1 V
8 0 V
7 -1 V
8 0 V
7 -1 V
8 0 V
7 -1 V
8 -1 V
7 0 V
8 -1 V
7 0 V
8 -1 V
7 0 V
8 -1 V
7 0 V
8 -1 V
7 0 V
8 -1 V
7 0 V
7 -1 V
8 0 V
7 -1 V
8 0 V
7 -1 V
8 -1 V
7 0 V
8 -1 V
7 0 V
8 -1 V
7 0 V
8 -1 V
7 0 V
8 -1 V
7 0 V
8 -1 V
7 0 V
8 -1 V
7 0 V
7 -1 V
8 0 V
7 -1 V
8 0 V
7 0 V
8 -1 V
7 0 V
8 -1 V
7 0 V
8 -1 V
7 0 V
8 -1 V
7 0 V
8 -1 V
7 0 V
8 -1 V
7 0 V
8 -1 V
7 0 V
7 -1 V
8 0 V
7 -1 V
8 0 V
7 0 V
8 -1 V
7 0 V
8 -1 V
7 0 V
8 -1 V
7 0 V
8 -1 V
7 0 V
8 0 V
7 -1 V
8 0 V
7 -1 V
8 0 V
7 -1 V
7 0 V
8 -1 V
7 0 V
8 0 V
7 -1 V
8 0 V
7 -1 V
8 0 V
7 -1 V
8 0 V
7 0 V
8 -1 V
7 0 V
8 -1 V
7 0 V
8 -1 V
7 0 V
8 0 V
7 -1 V
7 0 V
8 -1 V
7 0 V
8 0 V
stroke 6301 895 M
7 -1 V
8 0 V
7 -1 V
8 0 V
7 0 V
8 -1 V
7 0 V
8 -1 V
7 0 V
8 0 V
7 -1 V
8 0 V
7 -1 V
8 0 V
7 0 V
7 -1 V
8 0 V
7 -1 V
8 0 V
7 0 V
8 -1 V
7 0 V
8 0 V
7 -1 V
8 0 V
7 -1 V
8 0 V
7 0 V
8 -1 V
7 0 V
8 0 V
7 -1 V
8 0 V
7 -1 V
7 0 V
8 0 V
7 -1 V
8 0 V
7 0 V
8 -1 V
7 0 V
8 -1 V
7 0 V
8 0 V
7 -1 V
8 0 V
7 0 V
8 -1 V
7 0 V
8 0 V
7 -1 V
8 0 V
7 0 V
7 -1 V
8 0 V
7 -1 V
8 0 V
7 0 V
8 -1 V
7 0 V
8 0 V
7 -1 V
8 0 V
7 0 V
8 -1 V
7 0 V
8 0 V
7 -1 V
8 0 V
7 0 V
8 -1 V
7 0 V
stroke
1.000 UL
LTb
LCb setrgbcolor
860 4799 N
860 640 L
5979 0 V
0 4159 V
-5979 0 V
Z stroke
1.000 UP
1.000 UL
LTb
LCb setrgbcolor
stroke
grestore
end
showpage
  }}%
  \put(3849,140){\makebox(0,0){\strut{}$\frac{M_\rho^2}{\Lambda_\rho^2}$}}%
  \put(160,2719){%
  \special{ps: gsave currentpoint currentpoint translate
270 rotate neg exch neg exch translate}%
  \makebox(0,0){\strut{}$\zeta \left(2, \frac{M_\rho^2}{\Lambda_\rho^2} \right)$}%
  \special{ps: currentpoint grestore moveto}%
  }%
  \put(6839,440){\makebox(0,0){\strut{} 1}}%
  \put(6092,440){\makebox(0,0){\strut{} 0.9}}%
  \put(5344,440){\makebox(0,0){\strut{} 0.8}}%
  \put(4597,440){\makebox(0,0){\strut{} 0.7}}%
  \put(3849,440){\makebox(0,0){\strut{} 0.6}}%
  \put(3102,440){\makebox(0,0){\strut{} 0.5}}%
  \put(2355,440){\makebox(0,0){\strut{} 0.4}}%
  \put(1607,440){\makebox(0,0){\strut{} 0.3}}%
  \put(860,440){\makebox(0,0){\strut{} 0.2}}%
  \put(740,4799){\makebox(0,0)[r]{\strut{} 30}}%
  \put(740,4106){\makebox(0,0)[r]{\strut{} 25}}%
  \put(740,3413){\makebox(0,0)[r]{\strut{} 20}}%
  \put(740,2720){\makebox(0,0)[r]{\strut{} 15}}%
  \put(740,2026){\makebox(0,0)[r]{\strut{} 10}}%
  \put(740,1333){\makebox(0,0)[r]{\strut{} 5}}%
  \put(740,640){\makebox(0,0)[r]{\strut{} 0}}%
\end{picture}%
\endgroup
 

%% file: GdeRVFig3.tex
\begingroup%
\makeatletter%
\newcommand{\GNUPLOTspecial}{%
  \@sanitize\catcode`\%=14\relax\special}%
\setlength{\unitlength}{0.0500bp}%
\begin{picture}(7200,5040)(0,0)%
  {\GNUPLOTspecial{"
/gnudict 256 dict def
gnudict begin
%
%
/Color true def
/Blacktext true def
/Solid true def
/Dashlength 1 def
/Landscape false def
/Level1 false def
/Rounded false def
/ClipToBoundingBox false def
/SuppressPDFMark false def
/TransparentPatterns false def
/gnulinewidth 5.000 def
/userlinewidth gnulinewidth def
/Gamma 1.0 def
/BackgroundColor {-1.000 -1.000 -1.000} def
/vshift -66 def
/dl1 {
  10.0 Dashlength mul mul
  Rounded { currentlinewidth 0.75 mul sub dup 0 le { pop 0.01 } if } if
} def
/dl2 {
  10.0 Dashlength mul mul
  Rounded { currentlinewidth 0.75 mul add } if
} def
/hpt_ 31.5 def
/vpt_ 31.5 def
/hpt hpt_ def
/vpt vpt_ def
/doclip {
  ClipToBoundingBox {
    newpath 0 0 moveto 360 0 lineto 360 252 lineto 0 252 lineto closepath
    clip
  } if
} def
%
%
%
/M {moveto} bind def
/L {lineto} bind def
/R {rmoveto} bind def
/V {rlineto} bind def
/N {newpath moveto} bind def
/Z {closepath} bind def
/C {setrgbcolor} bind def
/f {rlineto fill} bind def
/g {setgray} bind def
/Gshow {show} def   
/vpt2 vpt 2 mul def
/hpt2 hpt 2 mul def
/Lshow {currentpoint stroke M 0 vshift R 
	Blacktext {gsave 0 setgray show grestore} {show} ifelse} def
/Rshow {currentpoint stroke M dup stringwidth pop neg vshift R
	Blacktext {gsave 0 setgray show grestore} {show} ifelse} def
/Cshow {currentpoint stroke M dup stringwidth pop -2 div vshift R 
	Blacktext {gsave 0 setgray show grestore} {show} ifelse} def
/UP {dup vpt_ mul /vpt exch def hpt_ mul /hpt exch def
  /hpt2 hpt 2 mul def /vpt2 vpt 2 mul def} def
/DL {Color {setrgbcolor Solid {pop []} if 0 setdash}
 {pop pop pop 0 setgray Solid {pop []} if 0 setdash} ifelse} def
/BL {stroke userlinewidth 2 mul setlinewidth
	Rounded {1 setlinejoin 1 setlinecap} if} def
/AL {stroke userlinewidth 2 div setlinewidth
	Rounded {1 setlinejoin 1 setlinecap} if} def
/UL {dup gnulinewidth mul /userlinewidth exch def
	dup 1 lt {pop 1} if 10 mul /udl exch def} def
/PL {stroke userlinewidth setlinewidth
	Rounded {1 setlinejoin 1 setlinecap} if} def
3.8 setmiterlimit
/LCw {1 1 1} def
/LCb {0 0 0} def
/LCa {0 0 0} def
/LC0 {1 0 0} def
/LC1 {0 1 0} def
/LC2 {0 0 1} def
/LC3 {1 0 1} def
/LC4 {0 1 1} def
/LC5 {1 1 0} def
/LC6 {0 0 0} def
/LC7 {1 0.3 0} def
/LC8 {0.5 0.5 0.5} def
/LTw {PL [] 1 setgray} def
/LTb {BL [] LCb DL} def
/LTa {AL [1 udl mul 2 udl mul] 0 setdash LCa setrgbcolor} def
/LT0 {PL [] LC0 DL} def
/LT1 {PL [4 dl1 2 dl2] LC1 DL} def
/LT2 {PL [2 dl1 3 dl2] LC2 DL} def
/LT3 {PL [1 dl1 1.5 dl2] LC3 DL} def
/LT4 {PL [6 dl1 2 dl2 1 dl1 2 dl2] LC4 DL} def
/LT5 {PL [3 dl1 3 dl2 1 dl1 3 dl2] LC5 DL} def
/LT6 {PL [2 dl1 2 dl2 2 dl1 6 dl2] LC6 DL} def
/LT7 {PL [1 dl1 2 dl2 6 dl1 2 dl2 1 dl1 2 dl2] LC7 DL} def
/LT8 {PL [2 dl1 2 dl2 2 dl1 2 dl2 2 dl1 2 dl2 2 dl1 4 dl2] LC8 DL} def
/Pnt {stroke [] 0 setdash gsave 1 setlinecap M 0 0 V stroke grestore} def
/Dia {stroke [] 0 setdash 2 copy vpt add M
  hpt neg vpt neg V hpt vpt neg V
  hpt vpt V hpt neg vpt V closepath stroke
  Pnt} def
/Pls {stroke [] 0 setdash vpt sub M 0 vpt2 V
  currentpoint stroke M
  hpt neg vpt neg R hpt2 0 V stroke
 } def
/Box {stroke [] 0 setdash 2 copy exch hpt sub exch vpt add M
  0 vpt2 neg V hpt2 0 V 0 vpt2 V
  hpt2 neg 0 V closepath stroke
  Pnt} def
/Crs {stroke [] 0 setdash exch hpt sub exch vpt add M
  hpt2 vpt2 neg V currentpoint stroke M
  hpt2 neg 0 R hpt2 vpt2 V stroke} def
/TriU {stroke [] 0 setdash 2 copy vpt 1.12 mul add M
  hpt neg vpt -1.62 mul V
  hpt 2 mul 0 V
  hpt neg vpt 1.62 mul V closepath stroke
  Pnt} def
/Star {2 copy Pls Crs} def
/BoxF {stroke [] 0 setdash exch hpt sub exch vpt add M
  0 vpt2 neg V hpt2 0 V 0 vpt2 V
  hpt2 neg 0 V closepath fill} def
/TriUF {stroke [] 0 setdash vpt 1.12 mul add M
  hpt neg vpt -1.62 mul V
  hpt 2 mul 0 V
  hpt neg vpt 1.62 mul V closepath fill} def
/TriD {stroke [] 0 setdash 2 copy vpt 1.12 mul sub M
  hpt neg vpt 1.62 mul V
  hpt 2 mul 0 V
  hpt neg vpt -1.62 mul V closepath stroke
  Pnt} def
/TriDF {stroke [] 0 setdash vpt 1.12 mul sub M
  hpt neg vpt 1.62 mul V
  hpt 2 mul 0 V
  hpt neg vpt -1.62 mul V closepath fill} def
/DiaF {stroke [] 0 setdash vpt add M
  hpt neg vpt neg V hpt vpt neg V
  hpt vpt V hpt neg vpt V closepath fill} def
/Pent {stroke [] 0 setdash 2 copy gsave
  translate 0 hpt M 4 {72 rotate 0 hpt L} repeat
  closepath stroke grestore Pnt} def
/PentF {stroke [] 0 setdash gsave
  translate 0 hpt M 4 {72 rotate 0 hpt L} repeat
  closepath fill grestore} def
/Circle {stroke [] 0 setdash 2 copy
  hpt 0 360 arc stroke Pnt} def
/CircleF {stroke [] 0 setdash hpt 0 360 arc fill} def
/C0 {BL [] 0 setdash 2 copy moveto vpt 90 450 arc} bind def
/C1 {BL [] 0 setdash 2 copy moveto
	2 copy vpt 0 90 arc closepath fill
	vpt 0 360 arc closepath} bind def
/C2 {BL [] 0 setdash 2 copy moveto
	2 copy vpt 90 180 arc closepath fill
	vpt 0 360 arc closepath} bind def
/C3 {BL [] 0 setdash 2 copy moveto
	2 copy vpt 0 180 arc closepath fill
	vpt 0 360 arc closepath} bind def
/C4 {BL [] 0 setdash 2 copy moveto
	2 copy vpt 180 270 arc closepath fill
	vpt 0 360 arc closepath} bind def
/C5 {BL [] 0 setdash 2 copy moveto
	2 copy vpt 0 90 arc
	2 copy moveto
	2 copy vpt 180 270 arc closepath fill
	vpt 0 360 arc} bind def
/C6 {BL [] 0 setdash 2 copy moveto
	2 copy vpt 90 270 arc closepath fill
	vpt 0 360 arc closepath} bind def
/C7 {BL [] 0 setdash 2 copy moveto
	2 copy vpt 0 270 arc closepath fill
	vpt 0 360 arc closepath} bind def
/C8 {BL [] 0 setdash 2 copy moveto
	2 copy vpt 270 360 arc closepath fill
	vpt 0 360 arc closepath} bind def
/C9 {BL [] 0 setdash 2 copy moveto
	2 copy vpt 270 450 arc closepath fill
	vpt 0 360 arc closepath} bind def
/C10 {BL [] 0 setdash 2 copy 2 copy moveto vpt 270 360 arc closepath fill
	2 copy moveto
	2 copy vpt 90 180 arc closepath fill
	vpt 0 360 arc closepath} bind def
/C11 {BL [] 0 setdash 2 copy moveto
	2 copy vpt 0 180 arc closepath fill
	2 copy moveto
	2 copy vpt 270 360 arc closepath fill
	vpt 0 360 arc closepath} bind def
/C12 {BL [] 0 setdash 2 copy moveto
	2 copy vpt 180 360 arc closepath fill
	vpt 0 360 arc closepath} bind def
/C13 {BL [] 0 setdash 2 copy moveto
	2 copy vpt 0 90 arc closepath fill
	2 copy moveto
	2 copy vpt 180 360 arc closepath fill
	vpt 0 360 arc closepath} bind def
/C14 {BL [] 0 setdash 2 copy moveto
	2 copy vpt 90 360 arc closepath fill
	vpt 0 360 arc} bind def
/C15 {BL [] 0 setdash 2 copy vpt 0 360 arc closepath fill
	vpt 0 360 arc closepath} bind def
/Rec {newpath 4 2 roll moveto 1 index 0 rlineto 0 exch rlineto
	neg 0 rlineto closepath} bind def
/Square {dup Rec} bind def
/Bsquare {vpt sub exch vpt sub exch vpt2 Square} bind def
/S0 {BL [] 0 setdash 2 copy moveto 0 vpt rlineto BL Bsquare} bind def
/S1 {BL [] 0 setdash 2 copy vpt Square fill Bsquare} bind def
/S2 {BL [] 0 setdash 2 copy exch vpt sub exch vpt Square fill Bsquare} bind def
/S3 {BL [] 0 setdash 2 copy exch vpt sub exch vpt2 vpt Rec fill Bsquare} bind def
/S4 {BL [] 0 setdash 2 copy exch vpt sub exch vpt sub vpt Square fill Bsquare} bind def
/S5 {BL [] 0 setdash 2 copy 2 copy vpt Square fill
	exch vpt sub exch vpt sub vpt Square fill Bsquare} bind def
/S6 {BL [] 0 setdash 2 copy exch vpt sub exch vpt sub vpt vpt2 Rec fill Bsquare} bind def
/S7 {BL [] 0 setdash 2 copy exch vpt sub exch vpt sub vpt vpt2 Rec fill
	2 copy vpt Square fill Bsquare} bind def
/S8 {BL [] 0 setdash 2 copy vpt sub vpt Square fill Bsquare} bind def
/S9 {BL [] 0 setdash 2 copy vpt sub vpt vpt2 Rec fill Bsquare} bind def
/S10 {BL [] 0 setdash 2 copy vpt sub vpt Square fill 2 copy exch vpt sub exch vpt Square fill
	Bsquare} bind def
/S11 {BL [] 0 setdash 2 copy vpt sub vpt Square fill 2 copy exch vpt sub exch vpt2 vpt Rec fill
	Bsquare} bind def
/S12 {BL [] 0 setdash 2 copy exch vpt sub exch vpt sub vpt2 vpt Rec fill Bsquare} bind def
/S13 {BL [] 0 setdash 2 copy exch vpt sub exch vpt sub vpt2 vpt Rec fill
	2 copy vpt Square fill Bsquare} bind def
/S14 {BL [] 0 setdash 2 copy exch vpt sub exch vpt sub vpt2 vpt Rec fill
	2 copy exch vpt sub exch vpt Square fill Bsquare} bind def
/S15 {BL [] 0 setdash 2 copy Bsquare fill Bsquare} bind def
/D0 {gsave translate 45 rotate 0 0 S0 stroke grestore} bind def
/D1 {gsave translate 45 rotate 0 0 S1 stroke grestore} bind def
/D2 {gsave translate 45 rotate 0 0 S2 stroke grestore} bind def
/D3 {gsave translate 45 rotate 0 0 S3 stroke grestore} bind def
/D4 {gsave translate 45 rotate 0 0 S4 stroke grestore} bind def
/D5 {gsave translate 45 rotate 0 0 S5 stroke grestore} bind def
/D6 {gsave translate 45 rotate 0 0 S6 stroke grestore} bind def
/D7 {gsave translate 45 rotate 0 0 S7 stroke grestore} bind def
/D8 {gsave translate 45 rotate 0 0 S8 stroke grestore} bind def
/D9 {gsave translate 45 rotate 0 0 S9 stroke grestore} bind def
/D10 {gsave translate 45 rotate 0 0 S10 stroke grestore} bind def
/D11 {gsave translate 45 rotate 0 0 S11 stroke grestore} bind def
/D12 {gsave translate 45 rotate 0 0 S12 stroke grestore} bind def
/D13 {gsave translate 45 rotate 0 0 S13 stroke grestore} bind def
/D14 {gsave translate 45 rotate 0 0 S14 stroke grestore} bind def
/D15 {gsave translate 45 rotate 0 0 S15 stroke grestore} bind def
/DiaE {stroke [] 0 setdash vpt add M
  hpt neg vpt neg V hpt vpt neg V
  hpt vpt V hpt neg vpt V closepath stroke} def
/BoxE {stroke [] 0 setdash exch hpt sub exch vpt add M
  0 vpt2 neg V hpt2 0 V 0 vpt2 V
  hpt2 neg 0 V closepath stroke} def
/TriUE {stroke [] 0 setdash vpt 1.12 mul add M
  hpt neg vpt -1.62 mul V
  hpt 2 mul 0 V
  hpt neg vpt 1.62 mul V closepath stroke} def
/TriDE {stroke [] 0 setdash vpt 1.12 mul sub M
  hpt neg vpt 1.62 mul V
  hpt 2 mul 0 V
  hpt neg vpt -1.62 mul V closepath stroke} def
/PentE {stroke [] 0 setdash gsave
  translate 0 hpt M 4 {72 rotate 0 hpt L} repeat
  closepath stroke grestore} def
/CircE {stroke [] 0 setdash 
  hpt 0 360 arc stroke} def
/Opaque {gsave closepath 1 setgray fill grestore 0 setgray closepath} def
/DiaW {stroke [] 0 setdash vpt add M
  hpt neg vpt neg V hpt vpt neg V
  hpt vpt V hpt neg vpt V Opaque stroke} def
/BoxW {stroke [] 0 setdash exch hpt sub exch vpt add M
  0 vpt2 neg V hpt2 0 V 0 vpt2 V
  hpt2 neg 0 V Opaque stroke} def
/TriUW {stroke [] 0 setdash vpt 1.12 mul add M
  hpt neg vpt -1.62 mul V
  hpt 2 mul 0 V
  hpt neg vpt 1.62 mul V Opaque stroke} def
/TriDW {stroke [] 0 setdash vpt 1.12 mul sub M
  hpt neg vpt 1.62 mul V
  hpt 2 mul 0 V
  hpt neg vpt -1.62 mul V Opaque stroke} def
/PentW {stroke [] 0 setdash gsave
  translate 0 hpt M 4 {72 rotate 0 hpt L} repeat
  Opaque stroke grestore} def
/CircW {stroke [] 0 setdash 
  hpt 0 360 arc Opaque stroke} def
/BoxFill {gsave Rec 1 setgray fill grestore} def
/Density {
  /Fillden exch def
  currentrgbcolor
  /ColB exch def /ColG exch def /ColR exch def
  /ColR ColR Fillden mul Fillden sub 1 add def
  /ColG ColG Fillden mul Fillden sub 1 add def
  /ColB ColB Fillden mul Fillden sub 1 add def
  ColR ColG ColB setrgbcolor} def
/BoxColFill {gsave Rec PolyFill} def
/PolyFill {gsave Density fill grestore grestore} def
/h {rlineto rlineto rlineto gsave closepath fill grestore} bind def
%
%
/PatternFill {gsave /PFa [ 9 2 roll ] def
  PFa 0 get PFa 2 get 2 div add PFa 1 get PFa 3 get 2 div add translate
  PFa 2 get -2 div PFa 3 get -2 div PFa 2 get PFa 3 get Rec
  TransparentPatterns {} {gsave 1 setgray fill grestore} ifelse
  clip
  currentlinewidth 0.5 mul setlinewidth
  /PFs PFa 2 get dup mul PFa 3 get dup mul add sqrt def
  0 0 M PFa 5 get rotate PFs -2 div dup translate
  0 1 PFs PFa 4 get div 1 add floor cvi
	{PFa 4 get mul 0 M 0 PFs V} for
  0 PFa 6 get ne {
	0 1 PFs PFa 4 get div 1 add floor cvi
	{PFa 4 get mul 0 2 1 roll M PFs 0 V} for
 } if
  stroke grestore} def
/languagelevel where
 {pop languagelevel} {1} ifelse
 2 lt
	{/InterpretLevel1 true def}
	{/InterpretLevel1 Level1 def}
 ifelse
%
%
/Level2PatternFill {
/Tile8x8 {/PaintType 2 /PatternType 1 /TilingType 1 /BBox [0 0 8 8] /XStep 8 /YStep 8}
	bind def
/KeepColor {currentrgbcolor [/Pattern /DeviceRGB] setcolorspace} bind def
<< Tile8x8
 /PaintProc {0.5 setlinewidth pop 0 0 M 8 8 L 0 8 M 8 0 L stroke} 
>> matrix makepattern
/Pat1 exch def
<< Tile8x8
 /PaintProc {0.5 setlinewidth pop 0 0 M 8 8 L 0 8 M 8 0 L stroke
	0 4 M 4 8 L 8 4 L 4 0 L 0 4 L stroke}
>> matrix makepattern
/Pat2 exch def
<< Tile8x8
 /PaintProc {0.5 setlinewidth pop 0 0 M 0 8 L
	8 8 L 8 0 L 0 0 L fill}
>> matrix makepattern
/Pat3 exch def
<< Tile8x8
 /PaintProc {0.5 setlinewidth pop -4 8 M 8 -4 L
	0 12 M 12 0 L stroke}
>> matrix makepattern
/Pat4 exch def
<< Tile8x8
 /PaintProc {0.5 setlinewidth pop -4 0 M 8 12 L
	0 -4 M 12 8 L stroke}
>> matrix makepattern
/Pat5 exch def
<< Tile8x8
 /PaintProc {0.5 setlinewidth pop -2 8 M 4 -4 L
	0 12 M 8 -4 L 4 12 M 10 0 L stroke}
>> matrix makepattern
/Pat6 exch def
<< Tile8x8
 /PaintProc {0.5 setlinewidth pop -2 0 M 4 12 L
	0 -4 M 8 12 L 4 -4 M 10 8 L stroke}
>> matrix makepattern
/Pat7 exch def
<< Tile8x8
 /PaintProc {0.5 setlinewidth pop 8 -2 M -4 4 L
	12 0 M -4 8 L 12 4 M 0 10 L stroke}
>> matrix makepattern
/Pat8 exch def
<< Tile8x8
 /PaintProc {0.5 setlinewidth pop 0 -2 M 12 4 L
	-4 0 M 12 8 L -4 4 M 8 10 L stroke}
>> matrix makepattern
/Pat9 exch def
/Pattern1 {PatternBgnd KeepColor Pat1 setpattern} bind def
/Pattern2 {PatternBgnd KeepColor Pat2 setpattern} bind def
/Pattern3 {PatternBgnd KeepColor Pat3 setpattern} bind def
/Pattern4 {PatternBgnd KeepColor Landscape {Pat5} {Pat4} ifelse setpattern} bind def
/Pattern5 {PatternBgnd KeepColor Landscape {Pat4} {Pat5} ifelse setpattern} bind def
/Pattern6 {PatternBgnd KeepColor Landscape {Pat9} {Pat6} ifelse setpattern} bind def
/Pattern7 {PatternBgnd KeepColor Landscape {Pat8} {Pat7} ifelse setpattern} bind def
} def
%
%
%
/PatternBgnd {
  TransparentPatterns {} {gsave 1 setgray fill grestore} ifelse
} def
%
%
/Level1PatternFill {
/Pattern1 {0.250 Density} bind def
/Pattern2 {0.500 Density} bind def
/Pattern3 {0.750 Density} bind def
/Pattern4 {0.125 Density} bind def
/Pattern5 {0.375 Density} bind def
/Pattern6 {0.625 Density} bind def
/Pattern7 {0.875 Density} bind def
} def
%
%
Level1 {Level1PatternFill} {Level2PatternFill} ifelse
/Symbol-Oblique /Symbol findfont [1 0 .167 1 0 0] makefont
dup length dict begin {1 index /FID eq {pop pop} {def} ifelse} forall
currentdict end definefont pop
Level1 SuppressPDFMark or 
{} {
/SDict 10 dict def
systemdict /pdfmark known not {
  userdict /pdfmark systemdict /cleartomark get put
} if
SDict begin [
  /Title (fig3.tex)
  /Subject (gnuplot plot)
  /Creator (gnuplot 4.6 patchlevel 4)
  /Author (gregoryvulvert)
  /CreationDate (Tue Dec  3 11:46:48 2013)
  /DOCINFO pdfmark
end
} ifelse
end
gnudict begin
gsave
doclip
0 0 translate
0.050 0.050 scale
0 setgray
newpath
BackgroundColor 0 lt 3 1 roll 0 lt exch 0 lt or or not {BackgroundColor C 1.000 0 0 7200.00 5040.00 BoxColFill} if
1.000 UL
LTb
LCb setrgbcolor
980 640 M
31 0 V
5828 0 R
-31 0 V
980 794 M
31 0 V
5828 0 R
-31 0 V
980 948 M
63 0 V
5796 0 R
-63 0 V
stroke
LTb
LCb setrgbcolor
980 1102 M
31 0 V
5828 0 R
-31 0 V
980 1256 M
31 0 V
5828 0 R
-31 0 V
980 1410 M
31 0 V
5828 0 R
-31 0 V
980 1564 M
63 0 V
5796 0 R
-63 0 V
stroke
LTb
LCb setrgbcolor
980 1718 M
31 0 V
5828 0 R
-31 0 V
980 1872 M
31 0 V
5828 0 R
-31 0 V
980 2026 M
31 0 V
5828 0 R
-31 0 V
980 2180 M
63 0 V
5796 0 R
-63 0 V
stroke
LTb
LCb setrgbcolor
980 2334 M
31 0 V
5828 0 R
-31 0 V
980 2488 M
31 0 V
5828 0 R
-31 0 V
980 2642 M
31 0 V
5828 0 R
-31 0 V
980 2797 M
63 0 V
5796 0 R
-63 0 V
stroke
LTb
LCb setrgbcolor
980 2951 M
31 0 V
5828 0 R
-31 0 V
980 3105 M
31 0 V
5828 0 R
-31 0 V
980 3259 M
31 0 V
5828 0 R
-31 0 V
980 3413 M
63 0 V
5796 0 R
-63 0 V
stroke
LTb
LCb setrgbcolor
980 3567 M
31 0 V
5828 0 R
-31 0 V
980 3721 M
31 0 V
5828 0 R
-31 0 V
980 3875 M
31 0 V
5828 0 R
-31 0 V
980 4029 M
63 0 V
5796 0 R
-63 0 V
stroke
LTb
LCb setrgbcolor
980 4183 M
31 0 V
5828 0 R
-31 0 V
980 4337 M
31 0 V
5828 0 R
-31 0 V
980 4491 M
31 0 V
5828 0 R
-31 0 V
980 4645 M
63 0 V
5796 0 R
-63 0 V
stroke
LTb
LCb setrgbcolor
980 4799 M
31 0 V
5828 0 R
-31 0 V
1029 640 M
0 31 V
0 4128 R
0 -31 V
1126 640 M
0 31 V
0 4128 R
0 -31 V
1224 640 M
0 63 V
0 4096 R
0 -63 V
stroke
LTb
LCb setrgbcolor
1322 640 M
0 31 V
0 4128 R
0 -31 V
1419 640 M
0 31 V
0 4128 R
0 -31 V
1517 640 M
0 31 V
0 4128 R
0 -31 V
1615 640 M
0 31 V
0 4128 R
0 -31 V
1712 640 M
0 63 V
0 4096 R
0 -63 V
stroke
LTb
LCb setrgbcolor
1810 640 M
0 31 V
0 4128 R
0 -31 V
1908 640 M
0 31 V
0 4128 R
0 -31 V
2005 640 M
0 31 V
0 4128 R
0 -31 V
2103 640 M
0 31 V
0 4128 R
0 -31 V
2201 640 M
0 63 V
0 4096 R
0 -63 V
stroke
LTb
LCb setrgbcolor
2298 640 M
0 31 V
0 4128 R
0 -31 V
2396 640 M
0 31 V
0 4128 R
0 -31 V
2494 640 M
0 31 V
0 4128 R
0 -31 V
2591 640 M
0 31 V
0 4128 R
0 -31 V
2689 640 M
0 63 V
0 4096 R
0 -63 V
stroke
LTb
LCb setrgbcolor
2787 640 M
0 31 V
0 4128 R
0 -31 V
2884 640 M
0 31 V
0 4128 R
0 -31 V
2982 640 M
0 31 V
0 4128 R
0 -31 V
3079 640 M
0 31 V
0 4128 R
0 -31 V
3177 640 M
0 63 V
0 4096 R
0 -63 V
stroke
LTb
LCb setrgbcolor
3275 640 M
0 31 V
0 4128 R
0 -31 V
3372 640 M
0 31 V
0 4128 R
0 -31 V
3470 640 M
0 31 V
0 4128 R
0 -31 V
3568 640 M
0 31 V
0 4128 R
0 -31 V
3665 640 M
0 63 V
0 4096 R
0 -63 V
stroke
LTb
LCb setrgbcolor
3763 640 M
0 31 V
0 4128 R
0 -31 V
3861 640 M
0 31 V
0 4128 R
0 -31 V
3958 640 M
0 31 V
0 4128 R
0 -31 V
4056 640 M
0 31 V
0 4128 R
0 -31 V
4154 640 M
0 63 V
0 4096 R
0 -63 V
stroke
LTb
LCb setrgbcolor
4251 640 M
0 31 V
0 4128 R
0 -31 V
4349 640 M
0 31 V
0 4128 R
0 -31 V
4447 640 M
0 31 V
0 4128 R
0 -31 V
4544 640 M
0 31 V
0 4128 R
0 -31 V
4642 640 M
0 63 V
0 4096 R
0 -63 V
stroke
LTb
LCb setrgbcolor
4740 640 M
0 31 V
0 4128 R
0 -31 V
4837 640 M
0 31 V
0 4128 R
0 -31 V
4935 640 M
0 31 V
0 4128 R
0 -31 V
5032 640 M
0 31 V
0 4128 R
0 -31 V
5130 640 M
0 63 V
0 4096 R
0 -63 V
stroke
LTb
LCb setrgbcolor
5228 640 M
0 31 V
0 4128 R
0 -31 V
5325 640 M
0 31 V
0 4128 R
0 -31 V
5423 640 M
0 31 V
0 4128 R
0 -31 V
5521 640 M
0 31 V
0 4128 R
0 -31 V
5618 640 M
0 63 V
0 4096 R
0 -63 V
stroke
LTb
LCb setrgbcolor
5716 640 M
0 31 V
0 4128 R
0 -31 V
5814 640 M
0 31 V
0 4128 R
0 -31 V
5911 640 M
0 31 V
0 4128 R
0 -31 V
6009 640 M
0 31 V
0 4128 R
0 -31 V
6107 640 M
0 63 V
0 4096 R
0 -63 V
stroke
LTb
LCb setrgbcolor
6204 640 M
0 31 V
0 4128 R
0 -31 V
6302 640 M
0 31 V
0 4128 R
0 -31 V
6400 640 M
0 31 V
0 4128 R
0 -31 V
6497 640 M
0 31 V
0 4128 R
0 -31 V
6595 640 M
0 63 V
0 4096 R
0 -63 V
stroke
LTb
LCb setrgbcolor
6693 640 M
0 31 V
0 4128 R
0 -31 V
6790 640 M
0 31 V
0 4128 R
0 -31 V
stroke
LTb
LCb setrgbcolor
980 4799 N
980 640 L
5859 0 V
0 4159 V
-5859 0 V
Z stroke
LCb setrgbcolor
LTb
LCb setrgbcolor
LTb
1.000 UP
1.000 UL
LTb
LCb setrgbcolor
2.200 UL
LT0
1.00 0.00 0.00 C 980 4612 M
2 -2 V
3 -2 V
2 -1 V
3 -1 V
2 -1 V
3 0 V
2 0 V
3 1 V
2 1 V
2 0 V
3 2 V
2 1 V
3 1 V
2 2 V
3 2 V
2 2 V
3 1 V
2 2 V
2 2 V
3 2 V
2 3 V
3 2 V
2 2 V
3 2 V
2 2 V
2 2 V
3 2 V
2 2 V
3 2 V
2 2 V
3 2 V
2 2 V
3 2 V
2 1 V
2 2 V
3 2 V
2 1 V
3 2 V
2 1 V
3 2 V
2 1 V
3 1 V
2 1 V
2 1 V
3 1 V
2 1 V
3 1 V
2 1 V
3 0 V
2 1 V
3 0 V
2 1 V
2 0 V
3 0 V
2 1 V
3 0 V
2 0 V
3 0 V
2 0 V
2 -1 V
3 0 V
2 0 V
3 -1 V
2 0 V
3 -1 V
2 0 V
3 -1 V
2 -1 V
2 -1 V
3 0 V
2 -1 V
3 -1 V
2 -2 V
3 -1 V
2 -1 V
3 -1 V
2 -2 V
2 -1 V
3 -2 V
2 -1 V
3 -2 V
2 -1 V
3 -2 V
2 -2 V
3 -2 V
2 -2 V
2 -2 V
3 -1 V
2 -3 V
3 -2 V
2 -2 V
3 -2 V
2 -2 V
2 -2 V
3 -3 V
2 -2 V
3 -3 V
2 -2 V
3 -3 V
2 -2 V
3 -3 V
2 -2 V
2 -3 V
3 -3 V
stroke 1234 4601 M
2 -2 V
3 -3 V
2 -3 V
3 -3 V
2 -3 V
3 -3 V
2 -3 V
2 -3 V
3 -3 V
2 -3 V
3 -3 V
2 -3 V
3 -3 V
2 -3 V
3 -3 V
2 -4 V
2 -3 V
3 -3 V
2 -4 V
3 -3 V
2 -3 V
3 -4 V
2 -3 V
2 -3 V
3 -4 V
2 -3 V
3 -4 V
2 -3 V
3 -4 V
2 -4 V
3 -3 V
2 -4 V
2 -3 V
3 -4 V
2 -4 V
3 -3 V
2 -4 V
3 -4 V
2 -4 V
3 -3 V
2 -4 V
2 -4 V
3 -4 V
2 -4 V
3 -3 V
2 -4 V
3 -4 V
2 -4 V
3 -4 V
2 -4 V
2 -4 V
3 -4 V
2 -4 V
3 -3 V
2 -4 V
3 -4 V
2 -4 V
2 -4 V
3 -4 V
2 -4 V
3 -4 V
2 -4 V
3 -4 V
2 -5 V
3 -4 V
2 -4 V
2 -4 V
3 -4 V
2 -4 V
3 -4 V
2 -4 V
3 -4 V
2 -4 V
3 -4 V
2 -5 V
2 -4 V
3 -4 V
2 -4 V
3 -4 V
2 -4 V
3 -4 V
2 -5 V
3 -4 V
2 -4 V
2 -4 V
3 -4 V
2 -4 V
3 -5 V
2 -4 V
3 -4 V
2 -4 V
2 -4 V
3 -5 V
2 -4 V
3 -4 V
2 -4 V
3 -4 V
2 -5 V
3 -4 V
2 -4 V
2 -4 V
3 -4 V
2 -5 V
3 -4 V
stroke 1488 4208 M
2 -4 V
3 -4 V
2 -4 V
3 -5 V
2 -4 V
2 -4 V
3 -4 V
2 -4 V
3 -5 V
2 -4 V
3 -4 V
2 -4 V
3 -5 V
2 -4 V
2 -4 V
3 -4 V
2 -4 V
3 -5 V
2 -4 V
3 -4 V
2 -4 V
2 -4 V
3 -5 V
2 -4 V
3 -4 V
2 -4 V
3 -4 V
2 -5 V
3 -4 V
2 -4 V
2 -4 V
3 -4 V
2 -5 V
3 -4 V
2 -4 V
3 -4 V
2 -4 V
3 -4 V
2 -5 V
2 -4 V
3 -4 V
2 -4 V
3 -4 V
2 -4 V
3 -5 V
2 -4 V
3 -4 V
2 -4 V
2 -4 V
3 -4 V
2 -4 V
3 -4 V
2 -5 V
3 -4 V
2 -4 V
2 -4 V
3 -4 V
2 -4 V
3 -4 V
2 -4 V
3 -4 V
2 -5 V
3 -4 V
2 -4 V
2 -4 V
3 -4 V
2 -4 V
3 -4 V
2 -4 V
3 -4 V
2 -4 V
3 -4 V
2 -4 V
2 -4 V
3 -4 V
2 -4 V
3 -4 V
2 -4 V
3 -4 V
2 -4 V
3 -4 V
2 -4 V
2 -4 V
3 -4 V
2 -4 V
3 -4 V
2 -4 V
3 -4 V
2 -4 V
2 -4 V
3 -4 V
2 -4 V
3 -4 V
2 -4 V
3 -4 V
2 -4 V
3 -4 V
2 -4 V
2 -4 V
3 -4 V
2 -4 V
3 -4 V
2 -3 V
3 -4 V
stroke 1742 3782 M
2 -4 V
3 -4 V
2 -4 V
2 -4 V
3 -4 V
2 -4 V
3 -4 V
2 -3 V
3 -4 V
2 -4 V
3 -4 V
2 -4 V
2 -4 V
3 -3 V
2 -4 V
3 -4 V
2 -4 V
3 -4 V
2 -3 V
2 -4 V
3 -4 V
2 -4 V
3 -4 V
2 -3 V
3 -4 V
2 -4 V
3 -4 V
2 -3 V
2 -4 V
3 -4 V
2 -4 V
3 -3 V
2 -4 V
3 -4 V
2 -4 V
3 -3 V
2 -4 V
2 -4 V
3 -3 V
2 -4 V
3 -4 V
2 -4 V
3 -3 V
2 -4 V
3 -4 V
2 -3 V
2 -4 V
3 -4 V
2 -3 V
3 -4 V
2 -3 V
3 -4 V
2 -4 V
2 -3 V
3 -4 V
2 -4 V
3 -3 V
2 -4 V
3 -3 V
2 -4 V
3 -4 V
2 -3 V
2 -4 V
3 -3 V
2 -4 V
3 -3 V
2 -4 V
3 -3 V
2 -4 V
3 -4 V
2 -3 V
2 -4 V
3 -3 V
2 -4 V
3 -3 V
2 -4 V
3 -3 V
2 -4 V
3 -3 V
2 -4 V
2 -3 V
3 -4 V
2 -3 V
3 -3 V
2 -4 V
3 -3 V
2 -4 V
3 -3 V
2 -4 V
2 -3 V
3 -4 V
2 -3 V
3 -3 V
2 -4 V
3 -3 V
2 -4 V
2 -3 V
3 -3 V
2 -4 V
3 -3 V
2 -4 V
3 -3 V
2 -3 V
3 -4 V
stroke 1996 3404 M
2 -3 V
2 -3 V
3 -4 V
2 -3 V
3 -3 V
2 -4 V
3 -3 V
2 -3 V
3 -4 V
2 -3 V
2 -3 V
3 -4 V
2 -3 V
3 -3 V
2 -3 V
3 -4 V
2 -3 V
3 -3 V
2 -4 V
2 -3 V
3 -3 V
2 -3 V
3 -4 V
2 -3 V
3 -3 V
2 -3 V
2 -3 V
3 -4 V
2 -3 V
3 -3 V
2 -3 V
3 -4 V
2 -3 V
3 -3 V
2 -3 V
2 -3 V
3 -3 V
2 -4 V
3 -3 V
2 -3 V
3 -3 V
2 -3 V
3 -3 V
2 -4 V
2 -3 V
3 -3 V
2 -3 V
3 -3 V
2 -3 V
3 -3 V
2 -3 V
3 -4 V
2 -3 V
2 -3 V
3 -3 V
2 -3 V
3 -3 V
2 -3 V
3 -3 V
2 -3 V
2 -3 V
3 -3 V
2 -3 V
3 -3 V
2 -3 V
3 -4 V
2 -3 V
3 -3 V
2 -3 V
2 -3 V
3 -3 V
2 -3 V
3 -3 V
2 -3 V
3 -3 V
2 -3 V
3 -3 V
2 -3 V
2 -3 V
3 -3 V
2 -3 V
3 -3 V
2 -3 V
3 -2 V
2 -3 V
3 -3 V
2 -3 V
2 -3 V
3 -3 V
2 -3 V
3 -3 V
2 -3 V
3 -3 V
2 -3 V
2 -3 V
3 -3 V
2 -3 V
3 -2 V
2 -3 V
3 -3 V
2 -3 V
3 -3 V
2 -3 V
2 -3 V
stroke 2249 3081 M
3 -3 V
2 -2 V
3 -3 V
2 -3 V
3 -3 V
2 -3 V
3 -3 V
2 -3 V
2 -2 V
3 -3 V
2 -3 V
3 -3 V
2 -3 V
3 -2 V
2 -3 V
3 -3 V
2 -3 V
2 -3 V
3 -2 V
2 -3 V
3 -3 V
2 -3 V
3 -3 V
2 -2 V
2 -3 V
3 -3 V
2 -3 V
3 -2 V
2 -3 V
3 -3 V
2 -3 V
3 -2 V
2 -3 V
2 -3 V
3 -3 V
2 -2 V
3 -3 V
2 -3 V
3 -2 V
2 -3 V
3 -3 V
2 -3 V
2 -2 V
3 -3 V
2 -3 V
3 -2 V
2 -3 V
3 -3 V
2 -2 V
3 -3 V
2 -3 V
2 -2 V
3 -3 V
2 -3 V
3 -2 V
2 -3 V
3 -3 V
2 -2 V
2 -3 V
3 -2 V
2 -3 V
3 -3 V
2 -2 V
3 -3 V
2 -3 V
3 -2 V
2 -3 V
2 -2 V
3 -3 V
2 -3 V
3 -2 V
2 -3 V
3 -2 V
2 -3 V
3 -2 V
2 -3 V
2 -3 V
3 -2 V
2 -3 V
3 -2 V
2 -3 V
3 -2 V
2 -3 V
3 -2 V
2 -3 V
2 -2 V
3 -3 V
2 -3 V
3 -2 V
2 -3 V
3 -2 V
2 -3 V
2 -2 V
3 -3 V
2 -2 V
3 -3 V
2 -2 V
3 -3 V
2 -2 V
3 -3 V
2 -2 V
2 -2 V
3 -3 V
2 -2 V
stroke 2503 2805 M
3 -3 V
2 -2 V
3 -3 V
2 -2 V
3 -3 V
2 -2 V
2 -3 V
3 -2 V
2 -2 V
3 -3 V
2 -2 V
3 -3 V
2 -2 V
3 -3 V
2 -2 V
2 -2 V
3 -3 V
2 -2 V
3 -3 V
2 -2 V
3 -2 V
2 -3 V
2 -2 V
3 -3 V
2 -2 V
3 -2 V
2 -3 V
3 -2 V
2 -2 V
3 -3 V
2 -2 V
2 -3 V
3 -2 V
2 -2 V
3 -3 V
2 -2 V
3 -2 V
2 -3 V
3 -2 V
2 -2 V
2 -3 V
3 -2 V
2 -2 V
3 -3 V
2 -2 V
3 -2 V
2 -3 V
3 -2 V
2 -2 V
2 -3 V
3 -2 V
2 -2 V
3 -2 V
2 -3 V
3 -2 V
2 -2 V
2 -3 V
3 -2 V
2 -2 V
3 -2 V
2 -3 V
3 -2 V
2 -2 V
3 -2 V
2 -3 V
2 -2 V
3 -2 V
2 -2 V
3 -3 V
2 -2 V
3 -2 V
2 -2 V
3 -3 V
2 -2 V
2 -2 V
3 -2 V
2 -3 V
3 -2 V
2 -2 V
3 -2 V
2 -2 V
3 -3 V
2 -2 V
2 -2 V
3 -2 V
2 -2 V
3 -3 V
2 -2 V
3 -2 V
2 -2 V
2 -2 V
3 -2 V
2 -3 V
3 -2 V
2 -2 V
3 -2 V
2 -2 V
3 -2 V
2 -3 V
2 -2 V
3 -2 V
2 -2 V
3 -2 V
2 -2 V
stroke 2757 2566 M
3 -2 V
2 -3 V
3 -2 V
2 -2 V
2 -2 V
3 -2 V
2 -2 V
3 -2 V
2 -2 V
3 -3 V
2 -2 V
3 -2 V
2 -2 V
2 -2 V
3 -2 V
2 -2 V
3 -2 V
2 -2 V
3 -2 V
2 -3 V
2 -2 V
3 -2 V
2 -2 V
3 -2 V
2 -2 V
3 -2 V
2 -2 V
3 -2 V
2 -2 V
2 -2 V
3 -2 V
2 -2 V
3 -2 V
2 -2 V
3 -3 V
2 -2 V
3 -2 V
2 -2 V
2 -2 V
3 -2 V
2 -2 V
3 -2 V
2 -2 V
3 -2 V
2 -2 V
3 -2 V
2 -2 V
2 -2 V
3 -2 V
2 -2 V
3 -2 V
2 -2 V
3 -2 V
2 -2 V
2 -2 V
3 -2 V
2 -2 V
3 -2 V
2 -2 V
3 -2 V
2 -2 V
3 -2 V
2 -2 V
2 -2 V
3 -2 V
2 -2 V
3 -2 V
2 -2 V
3 -2 V
2 -2 V
3 -1 V
2 -2 V
2 -2 V
3 -2 V
2 -2 V
3 -2 V
2 -2 V
3 -2 V
2 -2 V
3 -2 V
2 -2 V
2 -2 V
3 -2 V
2 -2 V
3 -2 V
2 -2 V
3 -1 V
2 -2 V
3 -2 V
2 -2 V
2 -2 V
3 -2 V
2 -2 V
3 -2 V
2 -2 V
3 -2 V
2 -1 V
2 -2 V
3 -2 V
2 -2 V
3 -2 V
2 -2 V
3 -2 V
2 -2 V
stroke 3011 2357 M
3 -2 V
2 -1 V
2 -2 V
3 -2 V
2 -2 V
3 -2 V
2 -2 V
3 -2 V
2 -1 V
3 -2 V
2 -2 V
2 -2 V
3 -2 V
2 -2 V
3 -2 V
2 -1 V
3 -2 V
2 -2 V
3 -2 V
2 -2 V
2 -2 V
3 -1 V
2 -2 V
3 -2 V
2 -2 V
3 -2 V
2 -1 V
2 -2 V
3 -2 V
2 -2 V
3 -2 V
2 -1 V
3 -2 V
2 -2 V
3 -2 V
2 -2 V
2 -1 V
3 -2 V
2 -2 V
3 -2 V
2 -2 V
3 -1 V
2 -2 V
3 -2 V
2 -2 V
2 -2 V
3 -1 V
2 -2 V
3 -2 V
2 -2 V
3 -1 V
2 -2 V
3 -2 V
2 -2 V
2 -1 V
3 -2 V
2 -2 V
3 -2 V
2 -1 V
3 -2 V
2 -2 V
2 -2 V
3 -1 V
2 -2 V
3 -2 V
2 -2 V
3 -1 V
2 -2 V
3 -2 V
2 -2 V
2 -1 V
3 -2 V
2 -2 V
3 -1 V
2 -2 V
3 -2 V
2 -2 V
3 -1 V
2 -2 V
2 -2 V
3 -1 V
2 -2 V
3 -2 V
2 -1 V
3 -2 V
2 -2 V
3 -2 V
2 -1 V
2 -2 V
3 -2 V
2 -1 V
3 -2 V
2 -2 V
3 -1 V
2 -2 V
2 -2 V
3 -1 V
2 -2 V
3 -2 V
2 -1 V
3 -2 V
2 -2 V
3 -1 V
2 -2 V
stroke 3265 2174 M
2 -2 V
3 -1 V
2 -2 V
3 -2 V
2 -1 V
3 -2 V
2 -2 V
3 -1 V
2 -2 V
2 -1 V
3 -2 V
2 -2 V
3 -1 V
2 -2 V
3 -2 V
2 -1 V
3 -2 V
2 -2 V
2 -1 V
3 -2 V
2 -1 V
3 -2 V
2 -2 V
3 -1 V
2 -2 V
2 -1 V
3 -2 V
2 -2 V
3 -1 V
2 -2 V
3 -2 V
2 -1 V
3 -2 V
2 -1 V
2 -2 V
3 -1 V
2 -2 V
3 -2 V
2 -1 V
3 -2 V
2 -1 V
3 -2 V
2 -2 V
2 -1 V
3 -2 V
2 -1 V
3 -2 V
2 -1 V
3 -2 V
2 -2 V
3 -1 V
2 -2 V
2 -1 V
3 -2 V
2 -1 V
3 -2 V
2 -2 V
3 -1 V
2 -2 V
2 -1 V
3 -2 V
2 -1 V
3 -2 V
2 -1 V
3 -2 V
2 -1 V
3 -2 V
2 -2 V
2 -1 V
3 -2 V
2 -1 V
3 -2 V
2 -1 V
3 -2 V
2 -1 V
3 -2 V
2 -1 V
2 -2 V
3 -1 V
2 -2 V
3 -1 V
2 -2 V
3 -1 V
2 -2 V
3 -1 V
2 -2 V
2 -1 V
3 -2 V
2 -1 V
3 -2 V
2 -1 V
3 -2 V
2 -1 V
2 -2 V
3 -1 V
2 -2 V
3 -1 V
2 -2 V
3 -1 V
2 -2 V
3 -1 V
2 -2 V
2 -1 V
3 -2 V
stroke 3519 2011 M
2 -1 V
3 -2 V
2 -1 V
3 -2 V
2 -1 V
3 -2 V
2 -1 V
2 -2 V
3 -1 V
2 -2 V
3 -1 V
2 -1 V
3 -2 V
2 -1 V
3 -2 V
2 -1 V
2 -2 V
3 -1 V
2 -2 V
3 -1 V
2 -2 V
3 -1 V
2 -1 V
2 -2 V
3 -1 V
2 -2 V
3 -1 V
2 -2 V
3 -1 V
2 -1 V
3 -2 V
2 -1 V
2 -2 V
3 -1 V
2 -2 V
3 -1 V
2 -1 V
3 -2 V
2 -1 V
3 -2 V
2 -1 V
2 -2 V
3 -1 V
2 -1 V
3 -2 V
2 -1 V
3 -2 V
2 -1 V
3 -1 V
2 -2 V
2 -1 V
3 -2 V
2 -1 V
3 -1 V
2 -2 V
3 -1 V
2 -2 V
2 -1 V
3 -1 V
2 -2 V
3 -1 V
2 -2 V
3 -1 V
2 -1 V
3 -2 V
2 -1 V
2 -1 V
3 -2 V
2 -1 V
3 -2 V
2 -1 V
3 -1 V
2 -2 V
3 -1 V
2 -1 V
2 -2 V
3 -1 V
2 -2 V
3 -1 V
2 -1 V
3 -2 V
2 -1 V
3 -1 V
2 -2 V
2 -1 V
3 -1 V
2 -2 V
3 -1 V
2 -1 V
3 -2 V
2 -1 V
2 -2 V
3 -1 V
2 -1 V
3 -2 V
2 -1 V
3 -1 V
2 -2 V
3 -1 V
2 -1 V
2 -2 V
3 -1 V
2 -1 V
3 -2 V
stroke 3773 1865 M
2 -1 V
3 -1 V
2 -2 V
3 -1 V
2 -1 V
2 -1 V
3 -2 V
2 -1 V
3 -1 V
2 -2 V
3 -1 V
2 -1 V
3 -2 V
2 -1 V
2 -1 V
3 -2 V
2 -1 V
3 -1 V
2 -2 V
3 -1 V
2 -1 V
2 -1 V
3 -2 V
2 -1 V
3 -1 V
2 -2 V
3 -1 V
2 -1 V
3 -2 V
2 -1 V
2 -1 V
3 -1 V
2 -2 V
3 -1 V
2 -1 V
3 -2 V
2 -1 V
3 -1 V
2 -1 V
2 -2 V
3 -1 V
2 -1 V
3 -1 V
2 -2 V
3 -1 V
2 -1 V
3 -2 V
2 -1 V
2 -1 V
3 -1 V
2 -2 V
3 -1 V
2 -1 V
3 -1 V
2 -2 V
3 -1 V
2 -1 V
2 -1 V
3 -2 V
2 -1 V
3 -1 V
2 -1 V
3 -2 V
2 -1 V
2 -1 V
3 -1 V
2 -2 V
3 -1 V
2 -1 V
3 -1 V
2 -2 V
3 -1 V
2 -1 V
2 -1 V
3 -2 V
2 -1 V
3 -1 V
2 -1 V
3 -1 V
2 -2 V
3 -1 V
2 -1 V
2 -1 V
3 -2 V
2 -1 V
3 -1 V
2 -1 V
3 -2 V
2 -1 V
3 -1 V
2 -1 V
2 -1 V
3 -2 V
2 -1 V
3 -1 V
2 -1 V
3 -1 V
2 -2 V
2 -1 V
3 -1 V
2 -1 V
3 -1 V
2 -2 V
3 -1 V
stroke 4027 1734 M
2 -1 V
3 -1 V
2 -1 V
2 -2 V
3 -1 V
2 -1 V
3 -1 V
2 -1 V
3 -2 V
2 -1 V
3 -1 V
2 -1 V
2 -1 V
3 -2 V
2 -1 V
3 -1 V
2 -1 V
3 -1 V
2 -1 V
3 -2 V
2 -1 V
2 -1 V
3 -1 V
2 -1 V
3 -2 V
2 -1 V
3 -1 V
2 -1 V
2 -1 V
3 -1 V
2 -2 V
3 -1 V
2 -1 V
3 -1 V
2 -1 V
3 -1 V
2 -2 V
2 -1 V
3 -1 V
2 -1 V
3 -1 V
2 -1 V
3 -1 V
2 -2 V
3 -1 V
2 -1 V
2 -1 V
3 -1 V
2 -1 V
3 -2 V
2 -1 V
3 -1 V
2 -1 V
3 -1 V
2 -1 V
2 -1 V
3 -2 V
2 -1 V
3 -1 V
2 -1 V
3 -1 V
2 -1 V
2 -1 V
3 -1 V
2 -2 V
3 -1 V
2 -1 V
3 -1 V
2 -1 V
3 -1 V
2 -1 V
2 -2 V
3 -1 V
2 -1 V
3 -1 V
2 -1 V
3 -1 V
2 -1 V
3 -1 V
2 -1 V
2 -2 V
3 -1 V
2 -1 V
3 -1 V
2 -1 V
3 -1 V
2 -1 V
3 -1 V
2 -1 V
2 -2 V
3 -1 V
2 -1 V
3 -1 V
2 -1 V
3 -1 V
2 -1 V
2 -1 V
3 -1 V
2 -1 V
3 -2 V
2 -1 V
3 -1 V
2 -1 V
3 -1 V
stroke 4281 1615 M
2 -1 V
2 -1 V
3 -1 V
2 -1 V
3 -1 V
2 -1 V
3 -2 V
2 -1 V
3 -1 V
2 -1 V
2 -1 V
3 -1 V
2 -1 V
3 -1 V
2 -1 V
3 -1 V
2 -1 V
3 -1 V
2 -2 V
2 -1 V
3 -1 V
2 -1 V
3 -1 V
2 -1 V
3 -1 V
2 -1 V
2 -1 V
3 -1 V
2 -1 V
3 -1 V
2 -1 V
3 -1 V
2 -1 V
3 -2 V
2 -1 V
2 -1 V
3 -1 V
2 -1 V
3 -1 V
2 -1 V
3 -1 V
2 -1 V
3 -1 V
2 -1 V
2 -1 V
3 -1 V
2 -1 V
3 -1 V
2 -1 V
3 -1 V
2 -1 V
3 -1 V
2 -2 V
2 -1 V
3 -1 V
2 -1 V
3 -1 V
2 -1 V
3 -1 V
2 -1 V
2 -1 V
3 -1 V
2 -1 V
3 -1 V
2 -1 V
3 -1 V
2 -1 V
3 -1 V
2 -1 V
2 -1 V
3 -1 V
2 -1 V
3 -1 V
2 -1 V
3 -1 V
2 -1 V
3 -1 V
2 -1 V
2 -1 V
3 -1 V
2 -1 V
3 -1 V
2 -1 V
3 -1 V
2 -1 V
3 -1 V
2 -1 V
2 -1 V
3 -1 V
2 -1 V
3 -2 V
2 -1 V
3 -1 V
2 -1 V
2 -1 V
3 -1 V
2 -1 V
3 -1 V
2 -1 V
3 -1 V
2 -1 V
3 -1 V
2 -1 V
2 -1 V
stroke 4534 1506 M
3 -1 V
2 -1 V
3 -1 V
2 -1 V
3 -1 V
2 -1 V
3 -1 V
2 -1 V
2 -1 V
3 -1 V
2 0 V
3 -1 V
2 -1 V
3 -1 V
2 -1 V
3 -1 V
2 -1 V
2 -1 V
3 -1 V
2 -1 V
3 -1 V
2 -1 V
3 -1 V
2 -1 V
2 -1 V
3 -1 V
2 -1 V
3 -1 V
2 -1 V
3 -1 V
2 -1 V
3 -1 V
2 -1 V
2 -1 V
3 -1 V
2 -1 V
3 -1 V
2 -1 V
3 -1 V
2 -1 V
3 -1 V
2 -1 V
2 -1 V
3 -1 V
2 -1 V
3 -1 V
2 -1 V
3 -1 V
2 -1 V
3 0 V
2 -1 V
2 -1 V
3 -1 V
2 -1 V
3 -1 V
2 -1 V
3 -1 V
2 -1 V
2 -1 V
3 -1 V
2 -1 V
3 -1 V
2 -1 V
3 -1 V
2 -1 V
3 -1 V
2 -1 V
2 -1 V
3 -1 V
2 0 V
3 -1 V
2 -1 V
3 -1 V
2 -1 V
3 -1 V
2 -1 V
2 -1 V
3 -1 V
2 -1 V
3 -1 V
2 -1 V
3 -1 V
2 -1 V
3 -1 V
2 -1 V
2 0 V
3 -1 V
2 -1 V
3 -1 V
2 -1 V
3 -1 V
2 -1 V
2 -1 V
3 -1 V
2 -1 V
3 -1 V
2 -1 V
3 -1 V
2 0 V
3 -1 V
2 -1 V
2 -1 V
3 -1 V
2 -1 V
stroke 4788 1407 M
3 -1 V
2 -1 V
3 -1 V
2 -1 V
3 -1 V
2 -1 V
2 0 V
3 -1 V
2 -1 V
3 -1 V
2 -1 V
3 -1 V
2 -1 V
3 -1 V
2 -1 V
2 -1 V
3 -1 V
2 0 V
3 -1 V
2 -1 V
3 -1 V
2 -1 V
2 -1 V
3 -1 V
2 -1 V
3 -1 V
2 -1 V
3 0 V
2 -1 V
3 -1 V
2 -1 V
2 -1 V
3 -1 V
2 -1 V
3 -1 V
2 -1 V
3 -1 V
2 0 V
3 -1 V
2 -1 V
2 -1 V
3 -1 V
2 -1 V
3 -1 V
2 -1 V
3 -1 V
2 0 V
3 -1 V
2 -1 V
2 -1 V
3 -1 V
2 -1 V
3 -1 V
2 -1 V
3 0 V
2 -1 V
3 -1 V
2 -1 V
2 -1 V
3 -1 V
2 -1 V
3 -1 V
2 0 V
3 -1 V
2 -1 V
2 -1 V
3 -1 V
2 -1 V
3 -1 V
2 -1 V
3 0 V
2 -1 V
3 -1 V
2 -1 V
2 -1 V
3 -1 V
2 -1 V
3 0 V
2 -1 V
3 -1 V
2 -1 V
3 -1 V
2 -1 V
2 -1 V
3 -1 V
2 0 V
3 -1 V
2 -1 V
3 -1 V
2 -1 V
3 -1 V
2 -1 V
2 0 V
3 -1 V
2 -1 V
3 -1 V
2 -1 V
3 -1 V
2 0 V
2 -1 V
3 -1 V
2 -1 V
3 -1 V
2 -1 V
stroke 5042 1315 M
3 -1 V
2 0 V
3 -1 V
2 -1 V
2 -1 V
3 -1 V
2 -1 V
3 0 V
2 -1 V
3 -1 V
2 -1 V
3 -1 V
2 -1 V
2 -1 V
3 0 V
2 -1 V
3 -1 V
2 -1 V
3 -1 V
2 -1 V
3 0 V
2 -1 V
2 -1 V
3 -1 V
2 -1 V
3 -1 V
2 0 V
3 -1 V
2 -1 V
2 -1 V
3 -1 V
2 0 V
3 -1 V
2 -1 V
3 -1 V
2 -1 V
3 -1 V
2 0 V
2 -1 V
3 -1 V
2 -1 V
3 -1 V
2 -1 V
3 0 V
2 -1 V
3 -1 V
2 -1 V
2 -1 V
3 0 V
2 -1 V
3 -1 V
2 -1 V
3 -1 V
2 -1 V
3 0 V
2 -1 V
2 -1 V
3 -1 V
2 -1 V
3 0 V
2 -1 V
3 -1 V
2 -1 V
2 -1 V
3 0 V
2 -1 V
3 -1 V
2 -1 V
3 -1 V
2 0 V
3 -1 V
2 -1 V
2 -1 V
3 -1 V
2 0 V
3 -1 V
2 -1 V
3 -1 V
2 -1 V
3 0 V
2 -1 V
2 -1 V
3 -1 V
2 -1 V
3 0 V
2 -1 V
3 -1 V
2 -1 V
3 -1 V
2 0 V
2 -1 V
3 -1 V
2 -1 V
3 -1 V
2 0 V
3 -1 V
2 -1 V
2 -1 V
3 0 V
2 -1 V
3 -1 V
2 -1 V
3 -1 V
2 0 V
stroke 5296 1231 M
3 -1 V
2 -1 V
2 -1 V
3 -1 V
2 0 V
3 -1 V
2 -1 V
3 -1 V
2 0 V
3 -1 V
2 -1 V
2 -1 V
3 -1 V
2 0 V
3 -1 V
2 -1 V
3 -1 V
2 0 V
3 -1 V
2 -1 V
2 -1 V
3 -1 V
2 0 V
3 -1 V
2 -1 V
3 -1 V
2 0 V
2 -1 V
3 -1 V
2 -1 V
3 0 V
2 -1 V
3 -1 V
2 -1 V
3 -1 V
2 0 V
2 -1 V
3 -1 V
2 -1 V
3 0 V
2 -1 V
3 -1 V
2 -1 V
3 0 V
2 -1 V
2 -1 V
3 -1 V
2 0 V
3 -1 V
2 -1 V
3 -1 V
2 0 V
3 -1 V
2 -1 V
2 -1 V
3 0 V
2 -1 V
3 -1 V
2 -1 V
3 0 V
2 -1 V
2 -1 V
3 -1 V
2 0 V
3 -1 V
2 -1 V
3 -1 V
2 0 V
3 -1 V
2 -1 V
2 -1 V
3 0 V
2 -1 V
3 -1 V
2 -1 V
3 0 V
2 -1 V
3 -1 V
2 -1 V
2 0 V
3 -1 V
2 -1 V
3 -1 V
2 0 V
3 -1 V
2 -1 V
3 -1 V
2 0 V
2 -1 V
3 -1 V
2 0 V
3 -1 V
2 -1 V
3 -1 V
2 0 V
2 -1 V
3 -1 V
2 -1 V
3 0 V
2 -1 V
3 -1 V
2 -1 V
3 0 V
2 -1 V
stroke 5550 1152 M
2 -1 V
3 0 V
2 -1 V
3 -1 V
2 -1 V
3 0 V
2 -1 V
3 -1 V
2 -1 V
2 0 V
3 -1 V
2 -1 V
3 0 V
2 -1 V
3 -1 V
2 -1 V
3 0 V
2 -1 V
2 -1 V
3 0 V
2 -1 V
3 -1 V
2 -1 V
3 0 V
2 -1 V
2 -1 V
3 0 V
2 -1 V
3 -1 V
2 -1 V
3 0 V
2 -1 V
3 -1 V
2 0 V
2 -1 V
3 -1 V
2 -1 V
3 0 V
2 -1 V
3 -1 V
2 0 V
3 -1 V
2 -1 V
2 -1 V
3 0 V
2 -1 V
3 -1 V
2 0 V
3 -1 V
2 -1 V
3 -1 V
2 0 V
2 -1 V
3 -1 V
2 0 V
3 -1 V
2 -1 V
3 0 V
2 -1 V
2 -1 V
3 -1 V
2 0 V
3 -1 V
2 -1 V
3 0 V
2 -1 V
3 -1 V
2 0 V
2 -1 V
3 -1 V
2 0 V
3 -1 V
2 -1 V
3 -1 V
2 0 V
3 -1 V
2 -1 V
2 0 V
3 -1 V
2 -1 V
3 0 V
2 -1 V
3 -1 V
2 0 V
3 -1 V
2 -1 V
2 -1 V
3 0 V
2 -1 V
3 -1 V
2 0 V
3 -1 V
2 -1 V
2 0 V
3 -1 V
2 -1 V
3 0 V
2 -1 V
3 -1 V
2 0 V
3 -1 V
2 -1 V
2 0 V
3 -1 V
stroke 5804 1079 M
2 -1 V
3 -1 V
2 0 V
3 -1 V
2 -1 V
3 0 V
2 -1 V
2 -1 V
3 0 V
2 -1 V
3 -1 V
2 0 V
3 -1 V
2 -1 V
3 0 V
2 -1 V
2 -1 V
3 0 V
2 -1 V
3 -1 V
2 0 V
3 -1 V
2 -1 V
3 0 V
2 -1 V
2 -1 V
3 0 V
2 -1 V
3 -1 V
2 0 V
3 -1 V
2 -1 V
2 0 V
3 -1 V
2 -1 V
3 0 V
2 -1 V
3 -1 V
2 0 V
3 -1 V
2 -1 V
2 0 V
3 -1 V
2 -1 V
3 0 V
2 -1 V
3 -1 V
2 0 V
3 -1 V
2 -1 V
2 0 V
3 -1 V
2 -1 V
3 0 V
2 -1 V
3 0 V
2 -1 V
3 -1 V
2 0 V
2 -1 V
3 -1 V
2 0 V
3 -1 V
2 -1 V
3 0 V
2 -1 V
2 -1 V
3 0 V
2 -1 V
3 -1 V
2 0 V
3 -1 V
2 -1 V
3 0 V
2 -1 V
2 -1 V
3 0 V
2 -1 V
3 0 V
2 -1 V
3 -1 V
2 0 V
3 -1 V
2 -1 V
2 0 V
3 -1 V
2 -1 V
3 0 V
2 -1 V
3 -1 V
2 0 V
3 -1 V
2 0 V
2 -1 V
3 -1 V
2 0 V
3 -1 V
2 -1 V
3 0 V
2 -1 V
2 -1 V
3 0 V
2 -1 V
3 0 V
stroke 6058 1011 M
2 -1 V
3 -1 V
2 0 V
3 -1 V
2 -1 V
2 0 V
3 -1 V
2 -1 V
3 0 V
2 -1 V
3 0 V
2 -1 V
3 -1 V
2 0 V
2 -1 V
3 -1 V
2 0 V
3 -1 V
2 0 V
3 -1 V
2 -1 V
3 0 V
2 -1 V
2 -1 V
3 0 V
2 -1 V
3 0 V
2 -1 V
3 -1 V
2 0 V
2 -1 V
3 -1 V
2 0 V
3 -1 V
2 0 V
3 -1 V
2 -1 V
3 0 V
2 -1 V
2 -1 V
3 0 V
2 -1 V
3 0 V
2 -1 V
3 -1 V
2 0 V
3 -1 V
2 0 V
2 -1 V
3 -1 V
2 0 V
3 -1 V
2 -1 V
3 0 V
2 -1 V
3 0 V
2 -1 V
2 -1 V
3 0 V
2 -1 V
3 0 V
2 -1 V
3 -1 V
2 0 V
2 -1 V
3 -1 V
2 0 V
3 -1 V
2 0 V
3 -1 V
2 -1 V
3 0 V
2 -1 V
2 0 V
3 -1 V
2 -1 V
3 0 V
2 -1 V
3 0 V
2 -1 V
3 -1 V
2 0 V
2 -1 V
3 0 V
2 -1 V
3 -1 V
2 0 V
3 -1 V
2 0 V
3 -1 V
2 -1 V
2 0 V
3 -1 V
2 0 V
3 -1 V
2 -1 V
3 0 V
2 -1 V
2 0 V
3 -1 V
2 -1 V
3 0 V
2 -1 V
3 0 V
stroke 6312 947 M
2 -1 V
3 -1 V
2 0 V
2 -1 V
3 0 V
2 -1 V
3 -1 V
2 0 V
3 -1 V
2 0 V
3 -1 V
2 -1 V
2 0 V
3 -1 V
2 0 V
3 -1 V
2 0 V
3 -1 V
2 -1 V
3 0 V
2 -1 V
2 0 V
3 -1 V
2 -1 V
3 0 V
2 -1 V
3 0 V
2 -1 V
2 -1 V
3 0 V
2 -1 V
3 0 V
2 -1 V
3 0 V
2 -1 V
3 -1 V
2 0 V
2 -1 V
3 0 V
2 -1 V
3 -1 V
2 0 V
3 -1 V
2 0 V
3 -1 V
2 0 V
2 -1 V
3 -1 V
2 0 V
3 -1 V
2 0 V
3 -1 V
2 0 V
3 -1 V
2 -1 V
2 0 V
3 -1 V
2 0 V
3 -1 V
2 -1 V
3 0 V
2 -1 V
2 0 V
3 -1 V
2 0 V
3 -1 V
2 -1 V
3 0 V
2 -1 V
3 0 V
2 -1 V
2 0 V
3 -1 V
2 -1 V
3 0 V
2 -1 V
3 0 V
2 -1 V
3 0 V
2 -1 V
2 0 V
3 -1 V
2 -1 V
3 0 V
2 -1 V
3 0 V
2 -1 V
3 0 V
2 -1 V
2 -1 V
3 0 V
2 -1 V
3 0 V
2 -1 V
3 0 V
2 -1 V
2 -1 V
3 0 V
2 -1 V
3 0 V
2 -1 V
3 0 V
2 -1 V
3 0 V
stroke 6566 887 M
2 -1 V
2 -1 V
3 0 V
2 -1 V
3 0 V
2 -1 V
3 0 V
2 -1 V
3 0 V
2 -1 V
2 -1 V
3 0 V
2 -1 V
3 0 V
2 -1 V
3 0 V
2 -1 V
3 0 V
2 -1 V
2 -1 V
3 0 V
2 -1 V
3 0 V
2 -1 V
3 0 V
2 -1 V
2 0 V
3 -1 V
2 -1 V
3 0 V
2 -1 V
3 0 V
2 -1 V
3 0 V
2 -1 V
2 0 V
3 -1 V
2 0 V
3 -1 V
2 -1 V
3 0 V
2 -1 V
3 0 V
2 -1 V
2 0 V
3 -1 V
2 0 V
3 -1 V
2 0 V
3 -1 V
2 -1 V
3 0 V
2 -1 V
2 0 V
3 -1 V
2 0 V
3 -1 V
2 0 V
3 -1 V
2 0 V
2 -1 V
3 -1 V
2 0 V
3 -1 V
2 0 V
3 -1 V
2 0 V
3 -1 V
2 0 V
2 -1 V
3 0 V
2 -1 V
3 0 V
2 -1 V
3 -1 V
2 0 V
3 -1 V
2 0 V
2 -1 V
3 0 V
2 -1 V
3 0 V
2 -1 V
3 0 V
2 -1 V
3 0 V
2 -1 V
2 0 V
3 -1 V
2 -1 V
3 0 V
2 -1 V
3 0 V
2 -1 V
2 0 V
3 -1 V
2 0 V
3 -1 V
2 0 V
3 -1 V
2 0 V
3 -1 V
2 0 V
2 -1 V
stroke 6819 830 M
3 0 V
2 -1 V
3 -1 V
2 0 V
3 -1 V
2 0 V
3 -1 V
2 0 V
stroke
1.000 UL
LTb
LCb setrgbcolor
980 4799 N
980 640 L
5859 0 V
0 4159 V
-5859 0 V
Z stroke
1.000 UP
1.000 UL
LTb
LCb setrgbcolor
stroke
grestore
end
showpage
  }}%
  \put(3909,140){\makebox(0,0){\strut{}$\frac{s_0}{\Lambda_\rho^2}$}}%
  \put(160,2719){%
  \special{ps: gsave currentpoint currentpoint translate
270 rotate neg exch neg exch translate}%
  \makebox(0,0){\strut{}$\zeta^{''} \left(2, \frac{s_0}{\Lambda_\rho^2} \right)$}%
  \special{ps: currentpoint grestore moveto}%
  }%
  \put(6595,440){\makebox(0,0){\strut{} 24}}%
  \put(6107,440){\makebox(0,0){\strut{} 22}}%
  \put(5618,440){\makebox(0,0){\strut{} 20}}%
  \put(5130,440){\makebox(0,0){\strut{} 18}}%
  \put(4642,440){\makebox(0,0){\strut{} 16}}%
  \put(4154,440){\makebox(0,0){\strut{} 14}}%
  \put(3665,440){\makebox(0,0){\strut{} 12}}%
  \put(3177,440){\makebox(0,0){\strut{} 10}}%
  \put(2689,440){\makebox(0,0){\strut{} 8}}%
  \put(2201,440){\makebox(0,0){\strut{} 6}}%
  \put(1712,440){\makebox(0,0){\strut{} 4}}%
  \put(1224,440){\makebox(0,0){\strut{} 2}}%
  \put(860,4645){\makebox(0,0)[r]{\strut{} 2}}%
  \put(860,4029){\makebox(0,0)[r]{\strut{} 1.8}}%
  \put(860,3413){\makebox(0,0)[r]{\strut{} 1.6}}%
  \put(860,2797){\makebox(0,0)[r]{\strut{} 1.4}}%
  \put(860,2180){\makebox(0,0)[r]{\strut{} 1.2}}%
  \put(860,1564){\makebox(0,0)[r]{\strut{} 1}}%
  \put(860,948){\makebox(0,0)[r]{\strut{} 0.8}}%
\end{picture}%
\endgroup
 